\documentclass[sigconf]{acmart}

\AtBeginDocument{%
  \providecommand\BibTeX{{%
    \normalfont B\kern-0.5em{\scshape i\kern-0.25em b}\kern-0.8em\TeX}}}

\copyrightyear{2021} 
\acmYear{2021}  
\setcopyright{acmcopyright}\acmConference[SC '21]{The International Conference for High Performance Computing, Networking, Storage and Analysis}{November 14--19, 2021}{St. Louis, MO, USA}
\acmBooktitle{The International Conference for High Performance Computing, Networking, Storage and Analysis (SC '21), November 14--19, 2021, St. Louis, MO, USA}
\acmPrice{15.00}
\acmDOI{10.1145/3458817.3476141}
\acmISBN{978-1-4503-8442-1/21/11}

\usepackage{amsmath,amsfonts}
\usepackage{algorithmicx}
\usepackage{color,soul}%For text background highlight 
\usepackage{textcomp}%For text background highlight 

\usepackage{booktabs} % For formal tables

\usepackage{amsmath}
\usepackage{rotating}
\usepackage{booktabs}
\usepackage{listings}
\usepackage{hhline,colortbl}

\usepackage[ruled,linesnumbered]{algorithm2e} % For algorithms

\usepackage{subfig}
\usepackage{multirow}
\usepackage{wrapfig}
\usepackage[font=small]{caption}

\startPage{1}
\usepackage{tikz}

\usepackage{setspace}
\newcommand*\circled[1]{\tikz[baseline=(char.base)]{
            \node[shape=circle,fill,inner sep=1pt] (char) {\footnotesize \textcolor{white}{#1}};}}

\newtheorem{myrule}{Rule}

 \usepackage{graphicx}   
 \usepackage{subfig}

 \newcommand{\ignore}[1]{}

\newcommand{\topk}{\textbf{\textit{Dr. Top-\textit{k}}}}

\newcommand{\update}[1]{\textcolor{black}{$^{\textrm{}}${#1}}}
\newcommand{\blue}[1]{\textcolor{black}{$^{\textrm{}}${#1}}}

\usepackage{multicol}
\usepackage{enumitem}

\begin{document}

\title{{\topk}: Delegate-Centric Top-\textit{k} on GPUs}

\author{\fontsize{10}{12}\selectfont Anil Gaihre}\affiliation{\fontsize{8}{10}\selectfont \institution{Stevens Institute of Technology}
\country{}}

\author{\fontsize{10}{12}\selectfont Da Zheng}
\affiliation{\fontsize{8}{10}\selectfont
	\institution{Johns Hopkins University}
	\country{}
}

\author{\fontsize{10}{12}\selectfont Scott Weitze}
\affiliation{\fontsize{8}{10}\selectfont
	\institution{Stevens Institute of Technology}
	\country{}
}

\author{\fontsize{10}{12}\selectfont Lingda Li}
\affiliation{\fontsize{8}{10}\selectfont
	\institution{Brookhaven National Laboratory}
	\country{}
}

\author{\fontsize{10}{12}\selectfont Shuaiwen Leon Song}
\affiliation{\fontsize{8}{10}\selectfont
	\institution{University of Sydney and UW Seattle}
	\country{}
}
	
\author{\fontsize{10}{12}\selectfont  Caiwen Ding}
\affiliation{\fontsize{8}{10}\selectfont
	\institution{ University of Connecticut}
	\country{}}

\author{\fontsize{10}{12}\selectfont  Xiaoye S. Li}
\affiliation{\fontsize{8}{10}\selectfont
	\institution{Lawrence Berkeley National Laboratory}
	\country{}}
	
\author{\fontsize{10}{12}\selectfont Hang Liu}
\affiliation{\fontsize{8}{10}\selectfont
	\institution{Stevens Institute of Technology}
	\country{}}

% \settopmatter{printacmref=true}

 \fancyhead{}
 
% \author{Anil Gaihre, Xiaoye S. Li, Lingda Li, Hang Liu, }
% \date{}

\begin{abstract}
% Recent top-$k$ computation projects explore the possibility of revising various sorting algorithms to answer top-$k$ queries on GPUs. These endeavors, unfortunately, perform significantly more work than needed. This paper introduces {\topk}, a \underline{D}elegate-cent\underline{r}ic \underline{top-$k$} system on GPUs that can reduce the top-$k$ workloads significantly. Particularly, {\topk} contains three major contributions: First, we introduce a comprehensive design of the delegate-centric concept, including maximum delegate, top-$k$ delegate-based filtering, and $\beta$ delegate mechanisms to help reduce the workload for top-$k$ by more than 99\%. Second, due to the difficulty and importance of deriving a proper subrange size in {\topk}, we perform a rigorous theoretical analysis, coupled with experimental validations to identify the desirable subrange size for {\topk}. Third, we introduce four key designs to enable fast multi-GPU {\topk} computation. Taken together, {\topk} constantly outperforms the state-of-the-art.
Recent top-$k$ computation efforts explore the possibility of revising various sorting algorithms to answer top-$k$ queries on GPUs. These endeavors, unfortunately, perform significantly more work than needed. This paper introduces {\topk}, a \underline{D}elegate-cent\underline{r}ic \underline{top-$k$} system on GPUs that can reduce the top-$k$ workloads significantly. Particularly, it contains three major contributions: First, we introduce a comprehensive design of the delegate-centric concept, including maximum delegate, delegate-based filtering, and $\beta$ delegate mechanisms to help reduce the workload for top-$k$ up to more than 99\%. Second, due to the difficulty and importance of deriving a proper subrange size, we perform a rigorous theoretical analysis, coupled with thorough experimental validations to identify the desirable subrange size. Third, we introduce four key system optimizations to enable fast multi-GPU top-$k$ computation. Taken together, this work constantly outperforms the state-of-the-art.
\end{abstract}

% %anil to put the page number in the document for ACM 2017 SIGCONF Template
% \settopmatter{printfolios=true}
%~anil to put the page number in the document for ACM 2017 SIGCONF Template

\maketitle
% \vspace{-.2in}
\section{Introduction}

% \fixme{Change min-heap to priority queue}

% \textbf{Problem Definition.} This paper targets the twin related query problems, that is, k-selection and top-k. Particularly, \textbf{k-selection} refers to finding/selecting the $k^{th}$ largest or smallest element from an input sequence. For simplicity, this paper only focuses on finding the $k^{th}$ largest element but the designs can also easily support the query for the $k^{th}$ smallest element. \textbf{Top-$k$} slightly complicates the query. That is, it not only find the $k^{th}$ largest but also the elements that are larger than this element. In other words, top-$k$ finds 1$^{st}$ - $k^{th}$ elements.

% It is important to note that existing efforts, e.g.,~\cite{alabi2012fast}, often first use $k$-selection to find the k$^{th}$ element, subsequently scan through the input vector again to arrive at the top-$k$ answer. However, some implementations like~\cite{shanbhag2018efficient} can directly extract the top-$k$ elements. 

% top-$k$ is an important routine for datacenter applications, top-\textit{k} is also important for our HPC community. For instance, the two of the most mainstream GPU programming frameworks Tensorflow~\cite{tensorflow,tensorflowRequest} and Arrayfire~\cite{malcolm2012arrayfire,arrayfireRequest} have feature requests to add a top-\textit{k} operator.

%The k selection is a well known problem in computer science. 
Formally, top-$k$ algorithms find the top $k$ elements from an input vector $V$. Here, the criteria could be the top $k$ largest or smallest, or any other conditions of interest. 
For simplicity, we assume the default criterion in this paper to be the top $k$ largest. $k$-selection algorithm slightly differs from the top-$k$ algorithm, as $k$-selection only identifies the $k^{th}$ largest element from $V$.
These two algorithms serve as building blocks for a variety of applications, such as, High Performance Computing (HPC)~\cite{tensorflowRequest,arrayfireRequest}, Information Retrieval (IR)~\cite{broder2003efficient,ding2011faster}, deep learning training~\cite{mxnet_topk,wang2020fft,wang2020ezlda}, big data~\cite{nguyen2015trustworthy,gaihre2018bitcoin,gaihre2019deanonymizing}, and data mining~\cite{malkov2018efficient,zheng2020pm,pandey2020c}.
% ~\cite{shneidman2005cost,das2009top,blumensath2010normalized,dai2009subspace,foucart2011hard,needell2009cosamp,blanchard2013gpu,ilyas2008survey}. 
%In particular, the applications ranges from digital signal processing algorithms, like greedy algorithms in compressed sensing in~\cite{blumensath2010normalized,dai2009subspace,foucart2011hard,needell2009cosamp,blanchard2013gpu} to running a common type of SQL query, for instance finding top 100 expensive items in a shopping website.  A survey on top-k~\cite{ilyas2008survey} describes various problem and efficient solutions for top-k.
%{Top-$k$ element can be selected naively by selecting the $k^{th}$ element after sorting the original vector but that does more work than necessary. 
A textbook implementation of top-$k$ exploits priority queue, i.e., min-heap. That is, a priority queue at the size of $k$ will slide through the input vector.
For each encountered element that is larger than the minimum from the priority queue, we substitute the minimum of the priority queue by this encountered element. Eventually, this priority queue captures the top-$k$ largest elements for the input vector $V$.
% \fixme{there should be a better way to describe min-heap which keeps top-k elements.}
% Collectively, this algorithm introduces $\mathcal{O}{(Nlogk)}$ time complexity.  
%However, it will suffers from thread divergence,  issues for large $k$ query~\cite{shanbhag2018efficient}. Sorting based $k$-selections radix top-$k$, bucket top-$k$~\cite{alabi2012fast} and bitonic top-$k$~\cite{shanbhag2018efficient} are the state-of-the-art tools on $k$-selection.
%}
% ~\cite{abadi2016tensorflow,heimel2013hardware,malcolm2012arrayfire,mostak2013overview,pirk2016voodoo,yuan2013yin,tensorflow}

Recently, the interest in deploying top-$k$ computation on GPUs has surged for three major reasons. First, GPUs offer superior processing power and memory throughput comparing to the other processing hardware~\cite{gokhale2008hardware,vetter2018extreme,li2018warp}. For instance, the most recent A100 GPU~\cite{a100_datasheet}
% \cite{a100,a100_datasheet}
features an astonishing {312} Tera Floating Point Operations Per Second (TFLOPS) computing capability and {2,039} GB/s memory throughput. Second, both the existing leading supercomputers~\cite{summit}  and future exascale ones (e.g., Aurora~\cite{aurara}, Frontier~\cite{frontier} and El Capitan~\cite{capitan}) use GPUs as the major computing resources. Third, the majority of the applications that exploit top-$k$, such as IR~\cite{liu2018griffin}, deep learning~\cite{abadi2016tensorflow}, data mining~\cite{zhao2020song,ren2020hm}\blue{, and database applications, e.g., PG Strom~\cite{pgstrom}, Ocelot~\cite{bress2014ocelot}, and MapD~\cite{root2016mapd}} are offloaded atop GPUs, deploying top-$k$ on GPUs could avoid copying data back and forth between GPU and CPU for top-$k$ computation. 
% Similarly, GPU based database applications like PG Strom~\cite{pgstrom}, Ocelot~\cite{bress2014ocelot}, and MapD~\cite{root2016mapd}, require to find the top-k elements in GPU. They are seeking for a GPU-based top-k system to avoid data transfers between CPU and GPU~\cite{shanbhag2018efficient}.

While priority queue-based top-$k$ is the most efficient approach for single- or multi-core systems~\cite{zois2019efficient}, it requires to maintain many local priority queues to expose massive parallelism to GPUs. Unfortunately, maintaining such many priority queues would experience expensive global synchronization overhead when merging these local priority queues into a final global one. Consequently, pertinent top-$k$ applications do not adopt priority queue-based top-$k$. \blue{Instead, they use sort-and-choose approach for top-$k$ computing on GPUs~\cite{stehle2017memory,guo2018memory,obeya2019theoretically,wang2020fft,bell2012thrust}. However, as shown in Figure~\ref{fig:DiffN_K_128Comparison}, the GPU-based sort-and-choose top-$k$~\cite{bell2012thrust} takes much longer time than GPU-based top-\textit{k} algorithms.}

Revising sorting algorithms to compute top-$k$ becomes a popular trend because, at most, only a subset of data needs to be sorted in the top-k problem. Along this direction, bitonic top-$k$~\cite{shanbhag2018efficient} presents a revised bitonic sort algorithm~\cite{ionescu1997optimizing} that focuses on the top-$k$ elements when merging 2$k$ elements together. Since this rudimentary design can only reduce the workload by half,~\cite{shanbhag2018efficient} further proposes to read \update{8$k$} elements and reduce it to $k$ while using GPU shared memory to cache the intermediate results. Due to the limited capacity of shared memory on GPUs, bitonic top-$k$ can only work for very small $k$ (i.e., \update{$k \leq 256$}). Another notable attempt~\cite{alabi2012fast} revises bucket sort by discarding all buckets that do not include the $k^{th}$ elements at each iteration, similarly for radix top-$k$. Despite these designs in~\cite{alabi2012fast} have the chance of reducing more workloads, they would suffer from unstable workload reductions (see Figure~\ref{sec:evaluation}).

% \vspace{.05in}
% {To reduce \textit{more} workload in a \textit{stable} manner, we introduce {\topk}, a delegate-centric system that partitions the input vector into subranges, extracts the delegate from each subrange, and uses the top-$k$ of the delegates to rapidly reduce the workload for top-$k$ computation over the input vector.} It is essential to note that the popular IR algorithm, i.e., Block Maximum WAND (BMW)~\cite{ding2011faster} also uses the delegate concept for search engine designs. In contrast, {\topk} {has a more comprehensive design and innovative usage} for the delegate concept. Taken together, {\topk} can help state-of-the-art top-$k$ algorithms to improve their performance significantly with the following three contributions: 
{To reduce \textit{more} workload in a \textit{stable} manner, we introduce {\topk}, a delegate-centric system that partitions the input vector into subranges, extracts the delegate from each subrange, and uses the top-$k$ of the delegates to rapidly reduce the workload for the overall top-$k$ computation on the input vector.} It is essential to note that the popular IR algorithm, i.e., Block Maximum WAND (BMW)~\cite{ding2011faster} also uses the delegate concept for search engine designs. In contrast, {\topk} {has a more comprehensive design and innovative usage} for the delegate concept. Taken together, it can help state-of-the-art top-$k$ algorithms to improve their performance significantly with the following three contributions:

\vspace{.05in}
First, we introduce a comprehensive delegate-centric design, which includes maximum delegate, top-$k$ delegate-based filtering, and $\beta$ delegate mechanisms to help reduce the workload for top-$k$ up to {more than 99\%}. 
Specifically, we (i) partition the input vector into a collection of subranges and extract the maximum delegate from each subrange to construct a delegate vector, and (ii) perform top-$k$ on the delegate vector. Since \textit{only those subranges whose maximum delegates belong to the top-$k$ of the delegate vector can contribute to the top-$k$ for the input vector}, we further (iii) concatenate those qualified subranges to construct a concatenated vector, and (iv) perform top-$k$ on the concatenated vector. To further reduce the workload for step (iv), we use the minimum of the top-$k$ of the delegate vector to filter out smaller elements from the qualified subranges. Not limited there, we extend the maximum delegate to $\beta$ delegate to reduce the workload for concatenation and second top-$k$. In particular, we will extract the top $\beta$ delegates, instead of merely the maximum, from each subrange.
% Afterward, we introduce a new rule -- \textit{If not all the top $\beta$ delegates from a subrange qualify for the top-$k$ of the delegate vector, the rest of the elements from this subrange will not contribute to the final top-$k$.} With this rule, we only concatenate subranges whose entire $\beta$ delegates are taken. 
Afterward, we introduce a new rule using which we only concatenate subranges whose entire $\beta$ delegates are taken.

\vspace{.05in}
Second, we deduce the optimal subrange size with both theoretical soundness and experimental validation. Note, a proper subrange size is crucial for {\topk} to achieve a good performance; on the one hand, a small subrange size would lead to too many subranges. 
In this context, the delegate vector construction and first top-$k$ would suffer from heavy workloads. On the other hand, when the subrange size is too large, we would have too few subranges. In this case, the majority of these subranges will be eligible for the second top-$k$. We hence skip too few subranges, leading to limited workload reduction for concatenation and second top-$k$. In Section~\ref{sec:alphaTune}, our theoretical analysis derives that the total time consumption of {\topk} is a {convex} function of subrange size, which we also verify in our experiment. 
We further extract the optimal subrange size for a wide range of $|V|$ and $k$.

\vspace{.05in}
Third, we deploy {\topk} atop multiple GPUs with four key system optimizations. First, we introduce a warp-centric delegate vector construction mechanism to achieve near-peak GPU global memory throughput. 
% \fixme{eval}
% Second, although that {\topk} can help all existing top-$k$ algorithms, we identify that the best {\topk} favors different top-$k$ algorithms when $k$ changes. We further introduce a flag-based strategy to avoid random memory access during in-place radix top-$k$. Third, we identify that during delegate vector construction, {\topk} suffers from low thread utilization and an exceeding usage of CUDA (an acronym for Compute Unified Device Architecture) shuffle instructions when $k$ becomes relatively large.
Second, although our delegate-centric design can help all existing top-$k$ algorithms, we identify that the best {\topk} assisted top-$k$ algorithm changes along with the climbing of $k$. We further introduce a flag-based strategy to avoid random memory access during in-place radix top-$k$. Third, we identify that delegate vector construction suffers from low thread utilization and an exceeding usage of CUDA (an acronym for Compute Unified Device Architecture) shuffle instructions when $k$ becomes relatively large.
% is not always fast which is also the cause of slow delegate vector construction. 
As a result, we introduce a novel \textit{coalesced load to shared memory and strided compute approach} to improve the thread utilization, as well as curb the usage of CUDA shuffle instructions. This optimization has reduced the delegate vector construction time consumption from {31.4 ms in Figure~\ref{fig:opt_concat_Uniform_With_Beta_split} to 9.5 ms in Figure~\ref{fig:SOK_decomposition_delegate_opt} for $k=2^{24}$ and $|V|=2^{30}$.} Finally, we scale top-$k$ across multiple GPUs to handle gigantic input vectors, and achieve sustained scalability. 
% \fixme{by \fixme{}$\times$.}

% , leading to significant speedup for delegate vector construction. 

% \vspace{.1in}
During evaluation, we notice that the recent top-$k$ efforts~\cite{alabi2012fast,shanbhag2018efficient} only test their systems on synthetic datasets, limiting the impacts of top-$k$. This paper hence builds a benchmark that contains three real-world applications, i.e., $k$-nearest neighbor search~\cite{jegou:inria-00514462},  website degree centrality~\cite{doo2014extracting}, and COVID fear related Twitter dataset~\cite{gupta2020global} for top-$k$. Our evaluation shows that {\topk} can outperform the state-of-the-art on both synthetic and real-world datasets. 

% As shown in Figures~\ref{fig:SOK_SpeedUp_different_distribution_Radix},~\ref{fig:SOK_SpeedUp_different_distribution_Bucket} and ~\ref{fig:SOK_SpeedUp_Bitonic}, we achieve upto 11.23$\times$, 9.23$\times$ and 12.23$\times$ speedups over radix top-$k$, bucket top-$k$ and bitonic top-$k$, respectively. Furthermore, the experimental results also anticipate the performance gains of {\topk} will climb with respect to the increase of the input vector.

% \vspace{.05in}
The remainder of this paper is organized as follows:
% Paper outline: Section~\ref{sec:background} discusses the background and related work which motivate overview in Section~\ref{sec:overview}. Section~\ref{sec:alg} presents the delegate-centric top-$k$ and compares {\topk} against BMW. Section~\ref{sec:system} deploys {\topk} on GPUs with GPU-specific optimizations. Section~\ref{sec:evaluation} evaluates {\topk} and we conclude in Section~\ref{sec:conclusion}.
Section~\ref{sec:background} discusses the background and related work which motivate the overview in Section~\ref{sec:overview}. Section~\ref{sec:alg} presents the delegate-centric top-$k$ design and compares it against BMW. Section~\ref{sec:system} deploys our top-$k$ on multi-GPU systems with GPU-specific optimizations. Section~\ref{sec:evaluation} evaluates {\topk} and we conclude in Section~\ref{sec:conclusion}.

% \begin{figure*}[t]
%  	\centering
%  	\includegraphics[scale=.48]{./figures/bucket_select.pdf}\vspace{-.15in} %1.png is the relative path of image
%  	\caption{bucket top-$k$ for an input vector with 16 elements (top-$2$). The shadow buckets are the ones of interest from one iteration to the next. 
% \vspace{-.28in}}.
%  	\label{fig:bucket_select} %  used for 
%  \end{figure*}

% \vspace{-.15in}
\section{Background and Related Work} \label{sec:background}

% This section presents the essential background and related work for {\topk}, i.e., GPU specification, and recent efforts in top-$k$.

% \vspace{-.2in}
\subsection{Graphics Processing Units}
% Originally designed for performing complex mathematical and geometric calculations that are necessary for graphical rendering, GPUs~\cite{6339609} have extended their application domains to scientific computing~\cite{6319192,chen2019tsm2,chabbi2013effective}, data analytics~\cite{wang2017gunrock}, machine learning~\cite{shi2018performance}, among many others. The rapid and steady enhancements in processing capability, memory capacity and data delivering ability from older to newer GPU generations have attracted numerous efforts to GPU computing. 
% We discuss the characteristics of GPUs that are closely related to the design of {\topk} with NVIDIA V100S GPU~\cite{voltagpu,volta_whitepaper} as an example.
% used in our experiments.

\textbf{Streaming processors and threads.} Designed with NVIDIA Volta architecture, V100S~\cite{voltagpu,volta_whitepaper} is powered by 80 streaming processors (SMs). Each SM is equipped with 64 CUDA cores, yielding a total of 5,120 cores running at 1.5 GHz. During execution, a GPU thread runs on one CUDA core, and an SM schedules a group of 32 consecutive threads known as warp in a Single Instruction Multiple Thread (SIMT) manner. Note, all the threads in a warp can use shuffle instructions to exchange data. A collection of consecutive warps further formulate a Cooperative Thread Arrays (CTAs) or a block. All the CTAs are called a grid. 

\textbf{Memory architecture.} V100S is equipped with 32 GB global memory with 1,134 GB/s as the peak throughput. All the SMs share an L2 cache of 6,144 KB. Each SM owns a private 96KB configurable shared memory, also used as the L1 cache. All the threads in a CTA can use shared memory to communicate with the help of the CUDA \_\_syncthreads() primitive. It is desirable to use shared memory to cache intermediate data because it is around one order of magnitude faster than the global memory~\cite{shanbhag2018efficient}.

% \vspace{-.15in}
\subsection{Related Work}\label{related}

This section discusses the closely related projects for {\topk} that includes priority queue-based top-$k$~\cite{shanbhag2018efficient}, sorting-based top-$k$~\cite{bell2012thrust,wang2020fft}, bucket top-$k$~\cite{alabi2012fast}, radix top-$k$~\cite{alabi2012fast} and bitonic top-$k$~\cite{shanbhag2018efficient}.

\textbf{Priority queue approach}.
A natural way to compute top-$k$ would be to maintain a priority queue that only keeps the top-$k$ elements while scanning through the input vector. While this idea is well-suited for single- or multi-threaded systems, implementing it on massively parallel GPUs remains elusive. Mainly, {a parallel implementation would involve the maintenance of many local priority queues and the eventual merging of these priority queues into a global one.}
This adds the challenge of frequent read and write and global synchronization across the threads when merging them. 
% Further, despite that the local priority queues could be maintained in shared memory to reduce the update overhead, the performance is likely to degrade with the increase of $k$ due to the concern of occupancy~\cite{luitjens2011cuda}.

\textbf{Sort-and-choose} is an alternative approach that is more friendly for parallel implementation.
Basically, we sort the input vector elements using sorting algorithms, like in THRUST~\cite{bell2012thrust} and choose the top-$k$ elements. But \textit{this implementation turns out to do more work than necessary.} At least, there is no need to sort the elements that are outside of the range of $1^{st}-k^{th}$ elements. Alabi et. al.~\cite{alabi2012fast} also show their top-$k$ algorithms, e.g., radix and bucket top-$k$ outperforms the sort-and-choose designs.

\textbf{Top-$k$ algorithms.} Bucket, radix, and bitonic top-$k$ are introduced to alleviate the aforementioned inefficiency problems faced by sorting. In contrast to their corresponding sort-and-choose approach, the top-$k$ algorithm distributes the input vector into different subranges, like a bucket in bucket top-$k$, and only focuses on the subrange that will lead to the $k^{th}$ element of the input vector.
Below we explain how these designs work with examples.

\textit{I. Bucket top-$k$} first obtains the $min$ and $max$ values from the input vector. Afterward, it divides this $min-max$ value range into several buckets, with each of which accounting for a disjoint equal value range. In the second step, this method scans through the input vector, puts each element into the corresponding bucket, and tracks the number of elements in each bucket. This way, one can easily figure out which bucket contains the $k^{th}$ elements. 
As mentioned earlier, top-$k$ operation discards the buckets that do not contain the $k^{th}$ element. This method continues until the bucket of interest only has one element, i.e., the $k^{th}$ one.

 \begin{figure}[t]
 	\centering
 	\includegraphics[scale=.259]{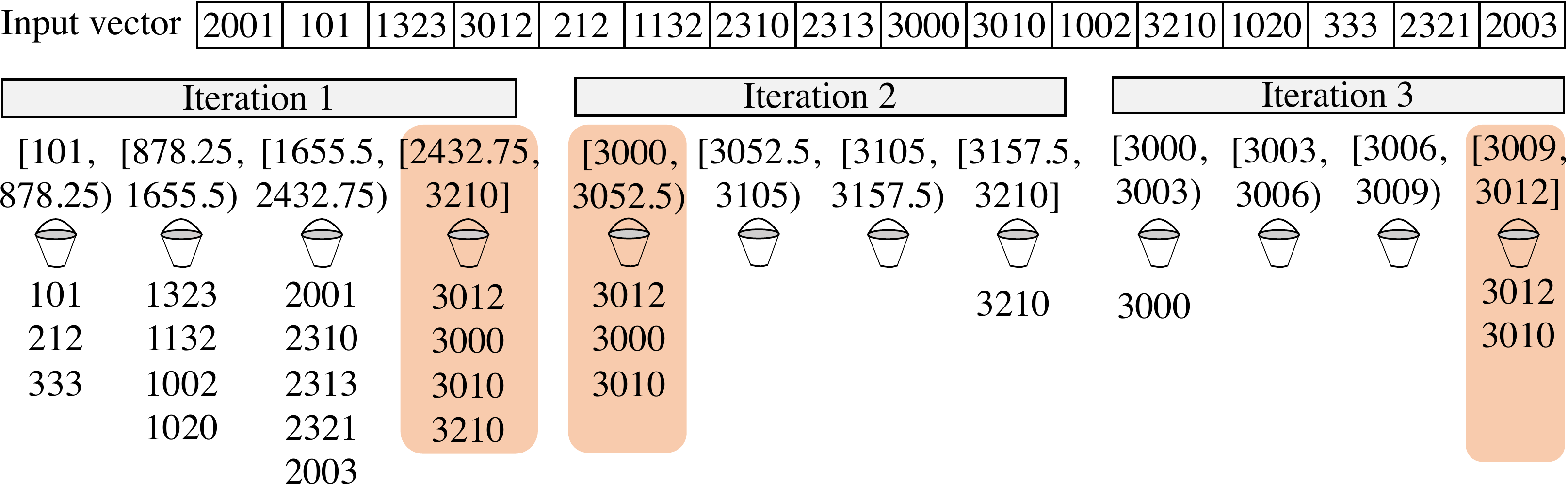}
%  	\vspace{-.15in}
 	\caption{Bucket top-2 computation for an input vector with 16 elements. The highlighted bucket is of interest at each iteration.
% \vspace{-.26in}
}
 	\label{fig:bucket_select} %  used for 
 \end{figure}
% \vspace{-.1in}

Figure~\ref{fig:bucket_select} exemplifies how bucket top-$k$ works for an input vector of 16 elements. We first derive the $min$ and $max$ as 101 and 3210, respectively. Therefore, we can divide this value range into four buckets, that is, {[101, 878.25), [878.25, 1655.5), [1655.5, 2432.75), [2432.75, 3210]}. Scanning through the entire input vector, one can obtain all elements that belong to each bucket as shown in iteration 1 of Figure~\ref{fig:bucket_select}. Since we are searching for the 2nd largest element from the input vector, our next iteration only focuses on the [2432.75, 3210] bucket, which is the largest bucket that contains four elements. Consequently, iteration 2 of Figure~\ref{fig:bucket_select} divides the [2432.75, 3210] value range into four buckets and scans through the elements from the [2432.75, 3210] bucket of iteration 1 to generate new element distributions in iteration 2 of Figure~\ref{fig:bucket_select}. 
While Figure~\ref{fig:bucket_select} only includes three iterations due to space constraints, this process is supposed to continue until the bucket of interest only contains one element.

% we find the top-2 elements. Limited by the space, we only draw three

% 3012 is the only element in the bucket of interest in iteration 4. %\td Can we use just "the figure" instead of referencing multiple times here?\td

%%%%\textit{II. Radix top-$k$} is similar to bucket top-$k$ but exploits the digits of each element to determine which bucket a value belongs to. The key is that \textit{the position of the bucket needs to indicate their order} so that we can derive the bucket of interest for the next iteration. Consequently, radix top-$k$ starts from the Most Significant Digit (MSD) to the Least Significant Digit (LSD). Following this manner, for instance, if we process 3 bits at one iteration, we will need eight buckets, that is, 000, 001, 010, 011, 100, 101, 110, 111. And all the elements from bucket `111' will be larger than those of `110'. Similarly for other buckets. At the end of each iteration, we will only focus on the bucket that contains the $k^{th}$ elements to proceed. 
%%%%%%%% Commented anil
\textit{II. Radix top-$k$} is similar to bucket top-$k$ but exploits the digits (i.e., radixes) of each element to determine which bucket a value belongs to. The key is that \textit{the position of the bucket needs to indicate their order} so that we can derive the bucket of interest for the next iteration. Consequently, radix top-$k$ starts from the Most Significant Digit (MSD) to the Least Significant Digit (LSD). Following this manner, for instance, if we process 3 bits at one iteration, we will need eight buckets, that is, 000, 001, 010, 011, 100, 101, 110, 111. And all the elements from bucket `111' are larger than those of `110'. {Similarly for other buckets.} At the end of each iteration, we only focus on the bucket that contains the $k^{th}$ elements to proceed. 
 
% \textit{III. Bitonic top-$k$}. Improving from the traditional bitonic {sorting} algorithm, bitonic top-$k$~\cite{shanbhag2018efficient} proposes to discard $k$ elements when combining two sorted $k$ sequences. Therefore, the workload is always reduced by half.
% Figure~\ref{fig:bitonic_select} demonstrates how bitonic top-2 behaves for the same input vector in Figure~\ref{fig:bucket_select}. Particularly, this algorithm sorts every two consecutive elements in the input vector, as shown in Iteration 1. Afterward, it merges the adjacent two sequences -- \{101, 2001\} and \{3012, 1323\} -- and gets the top-2 from these four elements, that is, \{2001, 3012\}, similarly for remaining sequences. This process continues until Iteration 3, where we obtain the final top-2, as \{3012, 3210\}.

\begin{figure}[t]
 	\centering
 	\includegraphics[scale=.45]{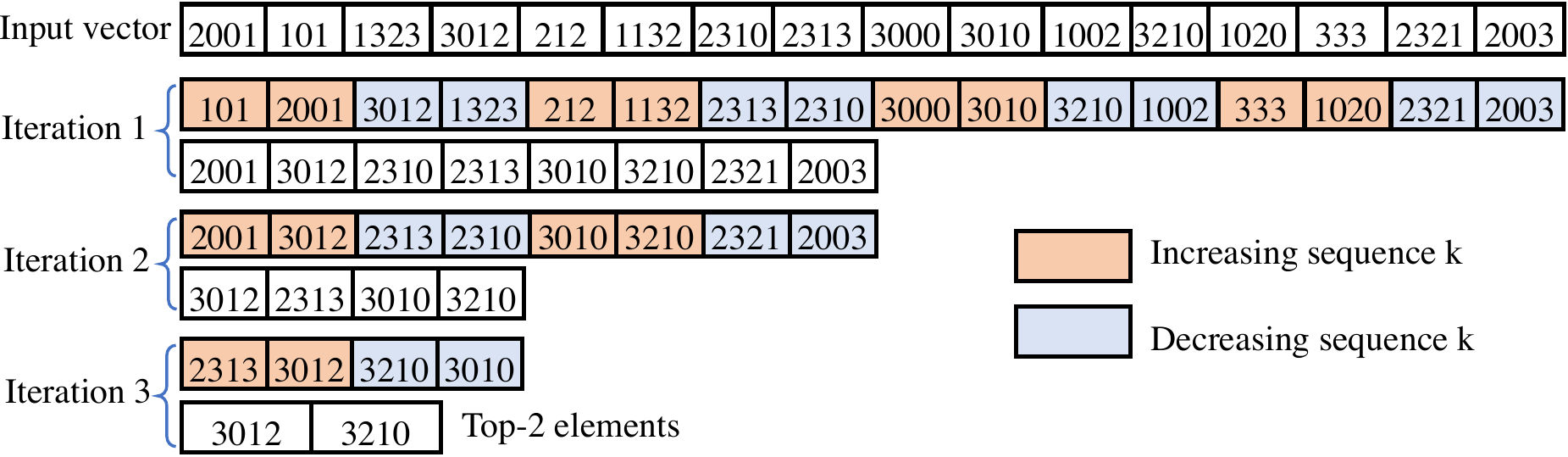} 
%  	\vspace{-.32in}
 	\caption{Bitonic top-2 on the same input vector as Figure~\ref{fig:bucket_select}.
%  	\vspace{-.25in}
 	}
 	\label{fig:bitonic_select} %  used for 
 \end{figure}

\textit{III. Bitonic top-$k$}. Improving from the traditional bitonic {sorting} algorithm, bitonic top-$k$~\cite{shanbhag2018efficient} proposes to discard $k$ elements by selecting the top-$k$ elements from a bitonic sequence of size 2$k$. Therefore, the workload is always reduced by half.
Figure~\ref{fig:bitonic_select} demonstrates how bitonic top-2 behaves for the same input vector in Figure~\ref{fig:bucket_select}. Particularly, this algorithm sorts every two consecutive elements in the input vector, as shown in Iteration 1. Afterward, it merges the adjacent two sequences -- \{101, 2001\} and \{3012, 1323\} -- and gets the top-2 from these four elements, that is, \{2001, 3012\}, similarly for remaining sequences. This process continues until Iteration 3, where we obtain the final top-2, as \{3012, 3210\}.

{Some of the other related projects worth mentioning are a GPU-based bucket sorting}~\cite{dehne2017parallel} that takes samples from different regions in the input vector to achieve a good workload balancing. {The work partitions the input vector into several subranges, performs a local sort in each subrange, and selects multiple samples from each subrange. These samples are collectively processed to guide data reordering on the original vector so that each bucket would end up with a similar amount of workloads.} 
{The top-$k$ at}~\cite{johnson2019billion} {performs a priority queue based k-selection algorithm in register memory in GPUs. As the registers per thread in the GPU are limited to a few numbers, similar to}~\cite{shanbhag2018efficient}{, the performance degrades for k $\geq$ 1024.} 
{A recent work}~\cite{ribizel2020parallel} {uses sampling to make bucket select more immune of skewed data distribution. Particularly, this work samples splitters from the original vector. Then, these splitters are sorted and used to assign bucket ranges. While this work tries to adjust the bucket boundaries in order to reduce workload in each level, \textit{our work directly reduces the original input vector for not only bucket top-$k$ but also other ones, e.g., radix and bitonic top-$k$ algorithms}.}

% \fixme{Add the related work~\cite{dehne2017parallel} concept here. Tell how the samples from different regions are used to create a well-balanced buckets for parallel implementation.$=>$ Tell not for workload reduction.}
% \fixme{Add also the work at~\cite{ribizel2020parallel} that uses concept of bucket select with sampling to make relatively distribution immune bucket select. Tell $\topk$ shows better speedup with respect to bucket select with sampling at any distribution. And {\topk} can be assist all the top-k selection including the work at~\cite{ribizel2020parallel}.}

%\vspace{-.2in}
\section{Challenges and Overview}\label{sec:overview}

% \blue{Figure~\ref{fig:inconsistency} presents the performance variation of {\topk} on three different distribution UD, Normal and Customized distribution.}

The state-of-the-art GPU-based top-$k$ designs, as shown in Figure~\ref{fig_arch}(a), directly work on the input vector when reducing the elements. This data reduction process continues until a desirable condition is met. Despite that such designs can outperform the traditional priority queue and sorting-based approaches, they still face the following two challenges:

\begin{figure}[t]
% \vspace{-.1in}
	\includegraphics[width=.49\textwidth]{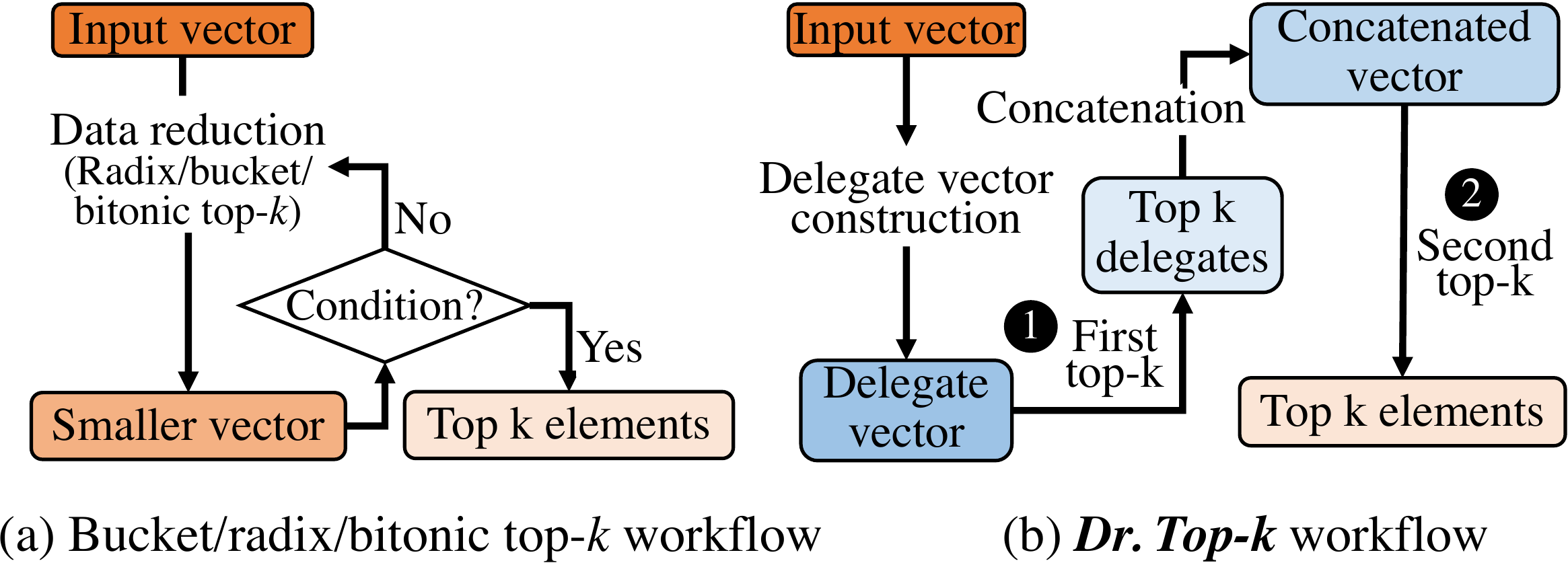}
% 	\vspace{-.2in}
	\caption {Workflow of (a) Bucket/radix/bitonic top-$k$ vs (b) {\topk}.
% 	\vspace{-.25in}
	}
	\label{fig_arch}
\end{figure}

\begin{figure}[ht]
	\centering
		\includegraphics[width=.47\textwidth]{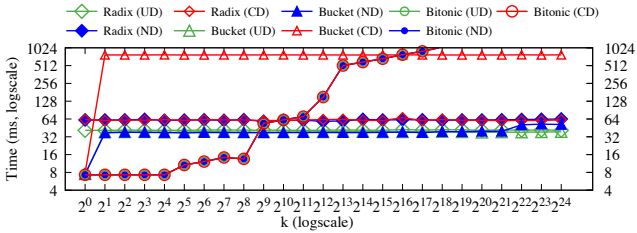}
% 	\vspace{-.15in} 
	\caption{\blue{The performance inconsistency of different top-$k$ versions on different distribution defined at Section~\ref{sec:evaluation}.}
% 	\vspace{-.2in}
	} 
	\label{fig:Inconsistency} 
% 	  \vspace{-.1in}	
\end{figure}

\begin{itemize}
    \item The performance of bucket and radix top-$k$ is unstable. That is, they are sensitive to the value distribution of the data. For instance, the radixes of interest might carry most of the elements from one iteration over to the next.  \blue{Figure~\ref{fig:Inconsistency} presents the performance variation of {\topk} on three different distributions Uniform (UD), Normal (ND) and Customized Distributions (CD), where the data distributions are rigorously defined in Section~\ref{sec:evaluation}. We observe both radix and bucket top-$k$~\cite{alabi2012fast} experience performance variations when changing data distributions. And bitonic top-k~\cite{shanbhag2018efficient} performs stably across different data distributions. 
    % This result is analogous to the analysis presented in~\cite{shanbhag2018efficient}.
    }
    \item While bitonic top-$k$ can stably reduce the workload, it only reduces the workload by half at one iteration. To further reduce the workload, bitonic top-$k$ requires tremendous shared memory to store the intermediate results. This is problematic for GPUs due to limited shared memory capacity. \blue{Figure~\ref{fig:bitonic_select} demonstrates how bitonic top-$k$ reduces the workload only by half at an iteration when it selects top-$2$ elements from each bitonic  
    % (increasing and decreasing) 
    sequence of length 4 at an iteration. For instance at iteration 1, from the first bitonic sequence of 4 elements \{101, 2001, 3012, 1323\} top-$2$ elements i.e. \{2001, 3013\} are selected to be written into new vector for next iteration. Similarly, remaining bitonic sequences in the vector go through same process. This leads the vector length to reduce from 16 to 8 at iteration 1.} 
\end{itemize}

{\topk}, as shown in Figure~\ref{fig_arch}(b), introduces the delegate-centric concept where top-$k$ computation only happens on 
%delegates or the subranges that contain the eligible delegates to overcome the challenges mentioned above, that is, {\topk} warrants both stable and larger workload reductions during top-$k$ computation. 
delegate and concatenated vectors which are small fraction of the original input vector. This warrants both stable and larger workload reductions during top-$k$ computation on {\topk}. 
(i) Our workload reduction is \textit{stable} regardless of the value distribution of the input vector. That said, for a given $k$ and $|V|$, the workload is determined (detailed in Section~\ref{sec:alphaTune}).  
% sizes of delegate and concatenated vector are the same, which indicates that the workload reduction is stable when the value distribution in $V$ changes.
% \fixme{this sentence is not clear.}
%This is because the input vector for these algorithms is significantly larger than the total size of delegate.
(ii) {\topk} on average reduces a greater portion of the workload, compared to top-$k$ algorithms such as bucket, radix, and bitonic top-$k$.
% (ii) {\topk} often reduces \textit{more} of the workload than bucket/radix/bitonic top-$k$ algorithms. 
This is because the total size of the delegate and concatenated vectors is smaller than the input vector, which is the input for the bucket/radix/bitonic top-$k$ algorithms. This is evident in Section~\ref{subsec:stat}.
% \fixme{maybe change the way to say this: you still need to go through the entire input vector when constructing delegates}
It should be noted that \textit{instead of being regarded as an alternative algorithm to the existing top-$k$ algorithms, {\topk} can help reduce workloads for all existing top-$k$ algorithms, including bucket, radix, and bitonic top-$k$ as long as we change the first and second top-$k$ (\circled{1}, and \circled{2}) into these algorithms. } 
% \fixme{describe how}

\begin{figure}[t]
% \vspace{-.1in}
	\includegraphics[width=.478\textwidth]{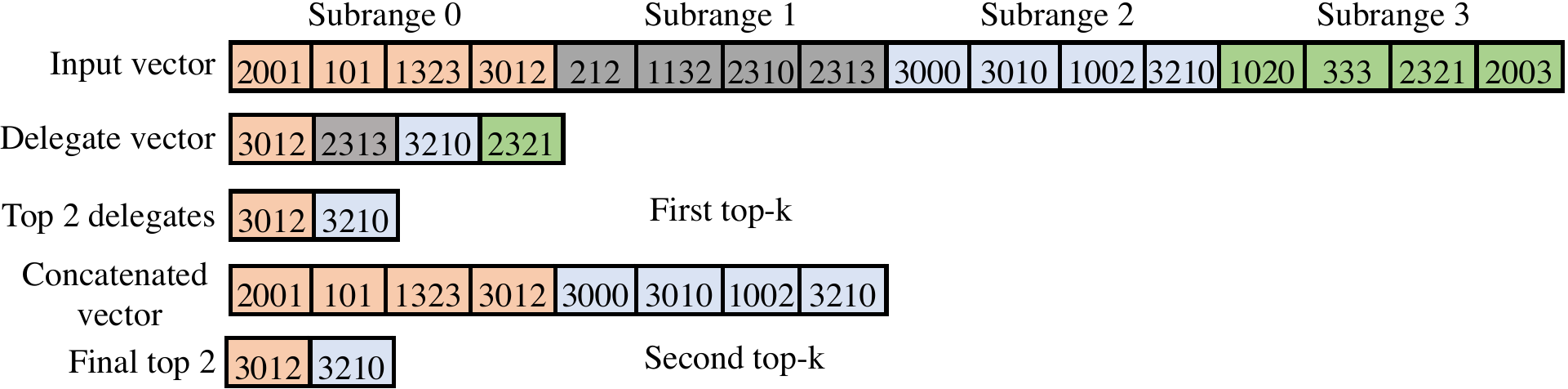}
% \hspace{-.1in}
	\caption {Maximum delegate-based top-2 computation for the same input vector $V$ in Figure~\ref{fig:bucket_select}. 
% 	\vspace{-.15in}
	}
	\label{fig:MaxSampling}
\end{figure}

\section{{\topk}: Delegate-Centric Top-\textit{k}}\label{sec:alg}
% \section{GPU Friendly Index Construction} \label{sec:system}

% This section discusses our comprehensive delegate-centric concept, which includes maximum delegate, delegate top-$k$ enabled filtering, and $\beta$ delegate. Further, we compare {\topk} against a popular and closely related IR algorithm, BMW.

% inspires the  {\topk} framework, as well as different phases of  {\topk}.
%Before discussing in details of  {\topk}, we introduce a principle that guides us through the implementation of  {\topk}.

\subsection{Maximum Delegate} \label{subsec:max}

\begin{myrule} \label{rule:max}
	%	Max Sampling: Sampling maximum element from a subrange: If the maximum from a range of numbers is not in the top-$k$, none of the numbers in this range will contribute to the top-$k$.	
% 	If the maximum $m_i$ from the subrange $S_i$ is not among the top-$k$ of the delegate vector, which contains the maximums of all the subranges, i.e., denoted as $D$, $S_i$ will not contribute to the top-$k$ for the input vector, i.e., $V$. 
	%Proposed Revision:
	{
	For a given vector $V$, $\exists D \in V$, such that $D$ is the delegate vector containing the maximum elements of all subranges $S_i$.
	If the maximum $m_i$ from $S_i$ is not among the top-$k$ elements in $D$, then $S_i$ will not contribute any elements to the top-$k$ of $V$.
	}
	%Note, $D$ is the set that contains all $m_i$.
\end{myrule}

Rule~\ref{rule:max} indicates that we can use the delegate of a subrange to decide whether to omit the entire subrange during top-$k$ computation. Figure~\ref{fig:MaxSampling} presents an example about how to use Rule~\ref{rule:max} to find the top $k$, (i.e., $k=2$) elements from the same input vector as Figure~\ref{fig:bucket_select}. Essentially, we first divide the input vector into four subranges, with each of which containing four elements. Second, we extract the delegate, i.e., maximum element from each subrange to formulate a delegate vector, i.e., \{3012, 2313, 3210, 2321\}. The first top-$k$ finds 3012 and 3210 as the top 2 elements from the delegate vector. This implies that only the subranges that contain 3012 and 3210 are qualified for concatenation. Therefore, we concatenate these two subranges and conduct the second top-$k$ on the concatenated vector -- \{2001, 101, 1323, 3012, 3000, 1002, 3210\}. Our second top-$k$ derives the final top-2 as \{3012, 3210\}.

\begin{figure}[hbt!]
	\centering
\includegraphics[width=0.47\textwidth]{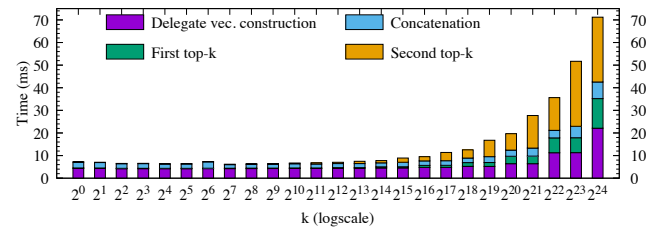}
% \vspace{-.1in}
% 	\caption{{\topk} assisted radix top-$k$ time consumption breakdown with respect to the increase of $k$ for $|V|=2^{30}$, where the elements in $V$ are generated by uniform distribution in the value range of 0 - $2^{32}$. This dataset is detailed in Section~\ref{sec:evaluation}. 
	\caption{{\topk} assisted radix top-$k$ time consumption breakdown with respect to the increase of $k$ for UD dataset on Section~\ref{sec:evaluation}.
% 	\vspace{-.30in}
	}
	\label{fig:SOK_decomposition_default} 
\end{figure}

% \textbf{Observation}.
Leveraging Rule~\ref{rule:max}, we implement the initial version of {\topk}.
Figure~\ref{fig:SOK_decomposition_default} demonstrates the time consumption breakdown of {\topk} accelerated radix top-$k$ for $|V|=2^{30}$ unsigned integers with $k$ ranging from $2^0$ to $2^{24}$. \textit{For $k\le2^{15}$, the time consumption delegate vector construction is $\sim$4.2 ms, which means we achieve 84\% of the peak throughput of the V100S GPU}, albeit delegate vector construction also performs additional shuffle instructions. When $k>2^{15}$, the time consumption of {\topk} also increases, which is reflected in all the four steps of {\topk}.

\subsection{Delegate Top-\textit{k} Enabled Filtering}\label{subsec:filter}

 \begin{figure}[t]
	\centering
\includegraphics[width=0.47\textwidth]{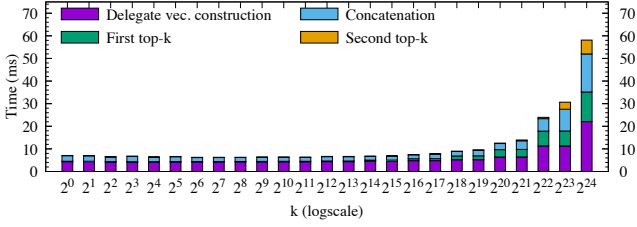}
% \vspace{-.1in}
% 	\caption{{\topk} with delegate top-$k$ enabled filtering for the $2^{30}$ elements that are uniformly generated, which is identical to the dataset used in Figure~\ref{fig:SOK_decomposition_default}.
	\caption{{\topk} with delegate top-$k$ enabled filtering for the UD dataset on Section~\ref{sec:evaluation}.
% 	\vspace{-.15in}
	}
	\label{fig:opt_concat_Uniform_noBeta_split} 
\end{figure}

Although the maximum delegate of a subrange is in the top-$k$ of the delegate vector, not all the elements in the qualified subrange will be eligible for the second top-$k$. This section uses the top-$k$ of the delegate vector to remove elements from the qualified subrange during the concatenation, leading to further workload reduction for the second top-$k$ through the following rule. 

\begin{myrule}\label{rule:filter}    
The $k^{th}$ element in the delegate vector is the minimum possible element the final $k^{th}$ element can become.
\end{myrule}

This rule can be derived as follows: the minimum of the top-$k$ of the input vector $V$ will be no less than that of the delegate vector $D$, i.e., $min(topk(D))\le min(topk(V))$. Therefore, only the elements that are larger than the $min(topk(D))$ are possible to get into $topk(V)$, hence are qualified for the concatenation. Here, $topk(V)$ denotes the top-$k$ elements in $V$, similarly for $topk(D)$.
We can use the example from Figure~\ref{fig:MaxSampling} to assist the understanding of this Rule~\ref{rule:filter}. %In the figure, we can observe that the size of the new vector is unnecessarily high.
Here, the minimum element from the top-2 of the delegate vector is 3012. Our prior {\topk} takes the entire subranges whose maximums are in the first top-$k$ into consideration. This is, in fact, wasteful. For instance, the elements that are smaller than 3012 in both subranges 0 and 2, that is, 2001, 101, 1323, 3000, 3010, and 1002, will never become one of the elements in the final top-2. Hence, none of them should be copied to the new concatenated vector. Eventually, the concatenated vector is merely \{3012, 3210\}.

To implement this delegate top-$k$ enabled filtering approach, we disseminate the minimum of the top-$k$ from the delegate vector across all threads. Afterward, the threads are dispatched to work on the qualified subranges identified by the first top-$k$. When performing scan on those qualified subranges, only the elements that are larger than the minimum of the top-$k$ of the delegate vector are stored in the concatenated vector. As the number of eligible elements from each subrange is unknown beforehand, each thread needs to use atomic operation~\cite{gaihre2019xbfs,gaihre2021gsofa} to obtain the position to store the eligible element.

Figure~\ref{fig:opt_concat_Uniform_noBeta_split} demonstrates the effectiveness of delegate top-$k$ enabled filtering for the same dataset in Figure~\ref{fig:SOK_decomposition_default}. 
% Note, this dataset follows a uniform distribution 
Comparing Figures~\ref{fig:opt_concat_Uniform_noBeta_split} and~\ref{fig:SOK_decomposition_default}, one can observe that the benefits of this optimization for the second top-$k$ is substantial, especially when $k\ge2^{19}$. Using $k=2^{24}$ as an example, we reduce the second top-$k$ time consumption from 28.7 ms in Figure~\ref{fig:SOK_decomposition_default} to 6.1 ms.
% Figure~\ref{fig:opt_concat_Uniform_noBeta_split} also demonstrates the overhead of filtering-based concatenation due to the extensive usage of atomic operation. Particularly, because an unknown number of elements from each subrange will contribute to the new vector, we rely on CUDA atomicAdd() to decide the correction position to store each eligible element. This introduces exceeding number of atomic operations. Comparing Figure~\ref{fig:SOK_decomposition_delegate_opt} and Figure~\ref{fig:opt_concat_Uniform_noBeta_split}, we can find the concatenation time increases from 25 ms to 28 ms when $k=2^{25}$ in spite of the high reduction in the memory writes.

  \begin{figure}[t]
 	% \centering
 	\includegraphics[width=.47\textwidth]{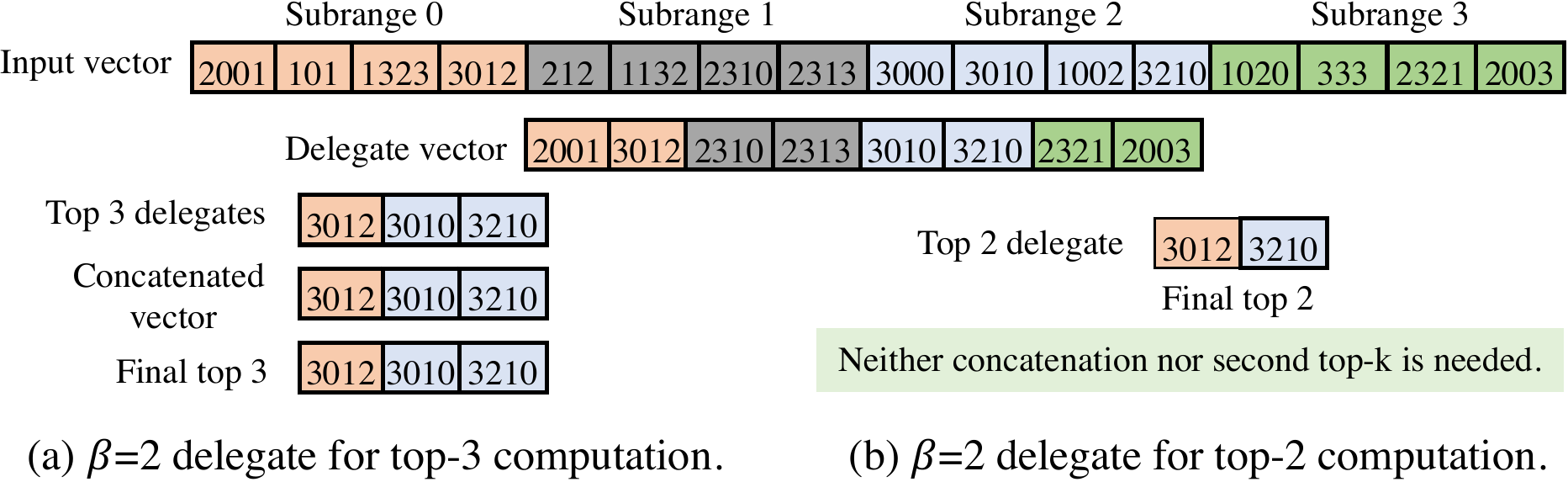}
%  	\vspace{-.1in}
 	\caption{Top-$k$ with $\beta$ delegate on the same input vector in Figure~\ref{fig:bucket_select}. Particularly, it shows different workload reductions for (a) $\beta=2$ delegate for top-3 query and (b) $\beta=2$ delegate for top-2 query.
%  	\vspace{-.1in}
 	}
 	\label{fig:betasampling}
 \end{figure}
 
\subsection{\texorpdfstring{$\beta$}{} Delegate} \label{subsec:betasampling}

While delegate top-$k$ enabled filtering can tremendously reduce the workload for second top-$k$, it still has two weaknesses that require further improvements: First, one might need to perform extensive atomic operations to build the concatenated vector. Second, we still need to scan through the qualified subranges to omit the elements that are smaller than the minimum of the top-$k$ of the delegate vector. 
% Figure~\ref{fig:opt_concat_Uniform_noBeta_split} also indicates that filtering-based concatenation uses atomic operations extensively, which experiences nontrivial overhead during concatenation. 
% Now we introduce $\beta$ delegate that allows {\topk} to safely avoid the entire subrange without scanning each element from the qualified subranges. Below, we formally introduce the $\beta$ delegate rule for this design.
Now we introduce $\beta$ delegate that allows {\topk} to safely avoid the entire subrange without scanning any elements which would be qualified for second top-$k$ if without $\beta$ delegates. Below, we formally introduce the $\beta$ delegate rule.

 \begin{myrule}\label{rule:beta}
 	In a subrange $S_i$, we select the top $\beta$ elements as $\beta$ delegates. {If not all of the $\beta$ delegates in $S_i$ would qualify as top-$k$ of the delegate vector $D$}, the rest of the elements from this subrange will not qualify for the top-$k$ of the input vector $V$. Note, $\beta\in N$ and $\beta>1$.
 \end{myrule}
 
 %Commented on Mar 17
%  \begin{myrule}\label{rule:beta}
%  	In a subrange $S_i$, we select the top $\beta$ elements as $\beta$ delegates. If this is not the case, then all of the $\beta$ delegates in $S_i$ would qualify as top-$k$ of the delegate vector $D$, the rest of the elements from this subrange will not qualify for the top-$k$ of the input vector $V$. Note, $\beta\in N$ and $\beta>1$.
%  \end{myrule}
 %Commented on Mar 17
 
 %Commented on Mar 17 2021
% \begin{proof}
% Assuming for subrange $S_i$, its $\beta$ delegates are $m^0_i, ..., m^{\beta-1}_i$, which are in descending order. That is, $m^0_i\ge m^1_i\ge ...\ge m^{\beta-1}_i$. During the top-$k$ computation on the delegate vector, i.e., $D$, because not all the $\beta$ delegates of $S_i$ are qualified for the top-$k$ of the delegate vector $D$, we have
% \begin{equation}\label{eq:beta}
% m^{\beta-1}_i<min(topk(D)).
% \end{equation}
% Since $min(topk(D))<min(topk(V))$, we obtain 
% \begin{equation}\label{eq:beta_2}
% m^{\beta-1}_i<min(topk(V)).
% \end{equation}
% Further, because the rest of the elements from $S_i$ are smaller than $m^{\beta-1}_i$, they are omitted from the top-$k$ of $V$.
% \end{proof} 
%Commented on Mar 17 2021

Figure~\ref{fig:betasampling} describes how to use Rule~\ref{rule:beta} to answer the top-3 and top-2 queries with $\beta =2$. In this case, our delegate vector contains two delegates from each subrange. 
% So instead of only using maximum delegate from each subranges, we obtain the top-$2$ elements to formulate the delegate vector. The first top-$k$ runs on this delegate vector to find the subranges for concatenation. 
For top-3 in Figure~\ref{fig:betasampling}(a), since we only take one delegate from subrange 0 and both delegates from subrange 2,
we obtain the concatenated vector as \{3012, 3010, 3210\}. Note, even the concatenated vector is the same as the top 3 delegates, we still need to scan through the entire subrange 3 to omit the ineligible elements. Finally, the second top-$k$ computes on the concatenated vector to figure out the final top-3. 
%
% Note, for ease of understanding, we intentionally omit the filtering optimization during concatenation in Figure~\ref{fig:betasampling}.
%
Figure~\ref{fig:betasampling}(b) presents a special benefit of $\beta$ delegate. That is, we might not need the concatenation and second top-$k$ computation. In this case, since the top-2 of the delegate vector does not take all the $\beta$ delegates from any subrange, Rule~\ref{rule:beta} suggests that neither concatenation nor second top-$k$ is necessary.

% , we find the the top-2 of the delegate vector as \{3012, 3210\}. This delegate top-2 does not take all the $\beta$ delegates from any subranges. According to Rule~\ref{rule:beta}, we do not need to concatenate any subranges for the second top-$k$ computation. Consequently, the first top-$k$ already gives the final top-$k$. 

Note, $\beta$ delegate will lead to more workloads for the first top-$k$ and delegate vector construction. To reduce the workload for the first top-$k$, we let the first top-$k$ skip the final iteration when locating the exact bucket or radix of interest. Because $\beta$ delegate and delegate top-$k$ enabled filtering can substantially reduce the workload for concatenation and second top-$k$, this skipping, which helps first top-$k$ noticeably, will lead to negligible performance drop for the subsequent concatenation and second top-$k$ steps. For performance improvements on delegate vector construction, Section~\ref{sub:sample_optimization} will introduce our novel optimizations shortly.
% \update{It is worth mentioning that skipping some of the iterations in first top-k can help reduce the first top-k time by negligible performance reduction in upcoming phases depending upon the distribution of the data.}

\begin{figure}[hbt!]
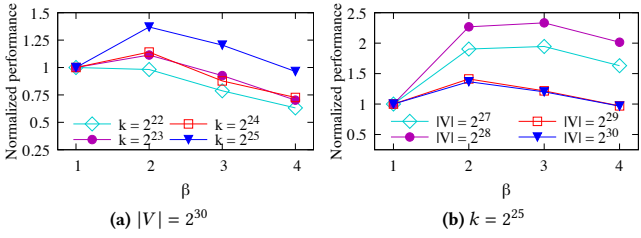

	\centering
% 		\vspace{-.2in}
	\begin{tabular}{cc}
		\hspace{-.1in}\subfloat[{$|V|=2^{30}$}]{\includegraphics[width=0.23\textwidth]{./figures/beta_tuning.pdf}} & 
		\hspace{-.05in}\subfloat[{$k=2^{25}$}]{\includegraphics[width=0.23\textwidth]{./figures/beta_tuningN.pdf}}\\
	\end{tabular}
% 	\vspace{-.15in}
	\caption{The performance dynamics with respect to the change of $\beta$ when (a) varying $k$ at $|V|=2^{30}$, and (b) varying $|V|$ at $k=2^{25}$. 
% \vspace{-.15in}
	}
	\label{fig:beta_tuning}
% 	\vspace{\shrink}
\end{figure}

An appropriate $\beta$ is important for {\topk}, our empirical study in
Figure~\ref{fig:beta_tuning} suggests that $\beta=2$ performs the best. For better visualization, we normalize the performance of various tests towards $\beta=1$. In Figure~\ref{fig:beta_tuning}(a), we find that $\beta=2$ is the desirable configuration which increases the performance up to 1.41 when $k = 2^{24}$ from $\beta=1$. Although Figure~\ref{fig:beta_tuning}(b) observes slightly better performance when $\beta=3$ for smaller $|V|=2^{29}$ and $2^{30}$, we find $\beta=2$ always yield good performance across both figures.

% \update{Figure~\ref{fig:beta_tuning}(a) presents the performance improvement of {\topk} over {\topk} without $\beta$ delegates (i.e. $\beta$ > 2). For large values of $k$ when the $\beta$ delegates are enabled, the performance of {\topk} is improved to a maximum of 1.41 for $k = 2^{24}$. A general trend of the curve concave downwards is observed signifying the degradation in performance for larger $\beta$. This  help in selection of optimum $\beta$ as 2 for our evaluations. Similarly, Figure~\ref{fig:beta_tuning}(b) shows a similar trend in the curves. We can observe that the smaller vectors |V| are getting more benefits from the $\beta$ delegates showing better performance in $\beta$=3 which is higher than optimum $\beta$=2 for the larger vectors of size $2^{29}$ and $2^{30}$.} In terms of the optimal $\beta$ value, {our experimental studies suggest that $\beta=2$ always yields the best performance for all $\beta$ delegate}. \fixme{Probably a figure}

  \begin{figure}[h]
	\centering
	\includegraphics[width=.47\textwidth]{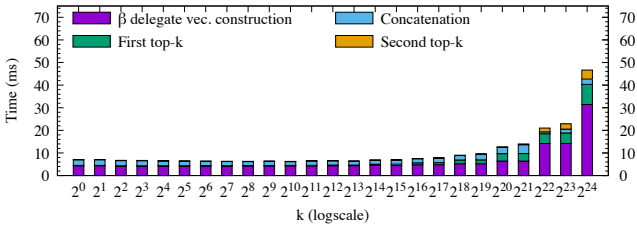}
% 	\vspace{-.15in} 
	\caption{{\topk} with $\beta$ delegate {and delegate top-$k$ enabled filtering optimization}  for UD dataset on Section~\ref{sec:evaluation}.
}
	\label{fig:opt_concat_Uniform_With_Beta_split} 
% 	  \vspace{-.2in}	
\end{figure}

%%%%%Commented Anil
% Figure~\ref{fig:opt_concat_Uniform_With_Beta_split} shows the time consumption breakdown of $\beta$ delegate optimization (including the filtering and delegate vector construction optimization). The most exciting news is that even for $k=2^{25}$, {\topk} still manages to only spend 28.38 ms to extract the top-$k$. Particularly, we spend 6.70 ms, 8.66 ms, 8.46 ms and 4.55 ms on delegate vector construction, first top-$k$, concatenation and second top-$k$, respectively. 
%%%%%Commented Anil
Figure~\ref{fig:opt_concat_Uniform_With_Beta_split} shows the time consumption breakdown of $\beta$ delegate optimization. Using k=$2^{24}$ as an example, although $\beta$ delegate spends 31.4 ms for delegate vector construction and {8.9} ms for first top-$k$, it reduces the time consumption for concatenation and second top-$k$ from 16.8 ms and 6.1 ms of Figure~\ref{fig:opt_concat_Uniform_noBeta_split} to 2.3 ms and 4 ms, respectively. Overall, we reduce the time consumption from 58 ms of Figure~\ref{fig:opt_concat_Uniform_noBeta_split} to {46.7} ms for $k=2^{24}$.

% \subsection{{\topk} vs Block Maximum WAND}
\subsection{Discussion: {\topk} vs BMW Algorithm}\label{subsec:bmw}

% \update{
% Information Retrieval (IR) system basically involves two phases. 1) the computation of scores to the documents based upon the query, and 2) top-$k$ selection of the documents based upon the score. 
% The two-level retrieval method WAND~\cite{broder2003efficient} performs top-k selection on two levels. The first level involves selection of candidate documents that are possible to be in the final top-$k$. The selection is done based upon approximate max scores of the terms in the query in the posting lists. A threshold $\lambda$ is maintained for the selection of the candidate documents. The second level in WAND performs full evaluation (compute total score) of candidate documents and gets top-$k$ documents with a minimum heap construction. The root of the minimum heap is used as $\lambda$ on the first level. This approach helps skip the full score computations of many documents while processing the query.} 

This section compares our {\topk} algorithm with a closely related IR algorithm\update{,} BMW~\cite{ding2011faster}\update{,} which is a variant of the popular Weak AND (WAND) algorithm~\cite{broder2003efficient}. Briefly, the BMW algorithm aims to find the top-$k$ most related documents for a query term. 

  \begin{figure}[t]
 	\centering
%  	\vspace{-.15in}
 	\includegraphics[width=.45\textwidth]{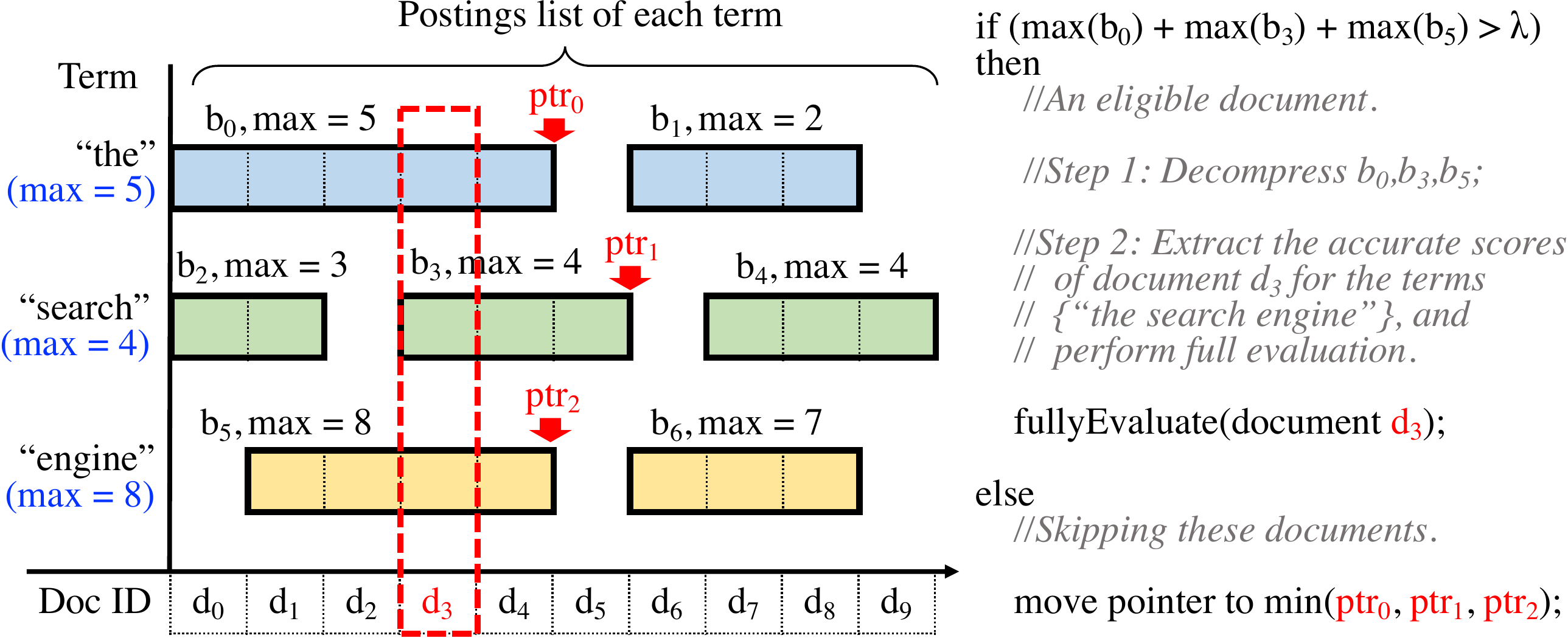} 
%  	\vspace{-.25in}
 	\caption{BMW algorithm for a query \{``the search engine''\}.
%  	Two level top-$k$ retrieval using the WAND operation. The documents (d1, d4 and d12) selected from the first level are fully evaluated second level with a maintenance of top-$k$ priority queue tree.
%  	\fixme{(1)Why d4 even if sum is already known why UB\_t is used? ==> For complex query terms, the upper bounds must be estimated since their posting lists are created dynamically during query evaluation. Estimation of UB results in requirement of balancing the false negatives and false positives (causes error). In overall there posting list won't have the document score per term. WAND doesn't score per document at its first level. It only uses UB\_t. (2)Point out the difference between the WAND and {\topk} based upon earlier sections.}
% \vspace{-.25in}
 	}
 	\label{fig:wand} 
 \end{figure}

Figure~\ref{fig:wand} presents an example for BMW algorithm. For clarity, we first describe the settings of our example: (i) there are ten documents, i.e., d$_0$ - d$_9$; (ii) the query contains three terms: ``the search engine'', and (iii) the score of a term in a document is the number of occurrence of the query term in that document. BMW first puts the documents that contain each term together, subsequently sorts them by the document ID and partitions them into blocks, e.g., the term ``the'' contains two blocks b$_0$ = \{d$_0$ - d$_4$\}, and b$_1$ = \{d$_6$ - d$_8$\}. For each block, BMW stores the maximum score, e.g., the maximum score of block b$_0$ is 5. Assuming BMW is working on document d$_3$, the right side of Figure~\ref{fig:wand} is the pseudocode of BMW. Specifically, BMW first evaluates whether the block maximums of the three blocks that contain document d$_3$ would be bigger than the threshold $\lambda$. If this condition is true, BMW will perform a full evaluation on document d$_3$ and move on to the next document d$_4$. Otherwise, BMW will skip all the documents that exactly share the same block maximum with document d$_3$.
In this context, using each block's maximum to estimate whether we should skip a document is similar to {\topk}, which uses the maximum of a subrange for delegate.

% finding the top-$k$ documents that have the most count of the search query words by using block maximum is similar to the top-$k$ algorithm that is performed in {\topk}. 
% \fixme{Hard to see why BMW is also computing top-k. May add a sentence to clarify the relationship of the problem you described above and top-k.}

\textbf{Distinction}.
% \fixme{Add more distinction of BMW with {\topk}.The reviewer wants conceptual difference not the performance difference.}
While BMW leverages the block maximum to skip computations when extracting the top-$k$ documents, {\topk} designs and exploits the delegate concept more comprehensively from three aspects. First, {\topk} introduces a delegate-centric processing concept while BMW still uses a regular element-centric concept. Here a regular element is a document. Particularly, {\topk} uses the delegate to decide whether an entire subrange (i.e., a block in BMW) is eligible or not. However, BMW processes one document at a time. Using document d$_3$ in Figure~\ref{fig:wand} as an example, even d$_3$ is qualified, BMW still needs to perform the eligibility check for d$_4$. {Further, for ineligible documents, BMW can only skip the documents that share the same or fewer terms than d$_3$}.
% , which is often rare according to~\cite{huang2017gpu}.
Second, we further introduce $\beta$ delegate to help remove more subranges and delegate top-$k$ enabled filtering to reduce some regular elements from the qualified subranges. Both of these designs are novel compared to BMW. Third, as we will discuss shortly in Section~\ref{sec:system}, {\topk} also includes subrange size tuning and GPU-aware optimizations, which are also novel compared to the BMW algorithm.

\section{{\topk} Implementation and Optimizations}\label{sec:system}

% This section unveils the implementation details of {\topk} on GPUs, $\alpha$ tuning and delegate vector construction optimizations.  %Algorithm~\ref{alg:delegate} sketches the pseudocode of this design.

\subsection{GPU-based {\topk} Design} 
\label{sec:gpu}

% \begin{algorithm}[h]
% 	\caption{\small Warp-centric delegate vector construction}
% 	\label{alg:SOK}
% 	\SetKwProg{generate}{Function {}}{}{end}
% {\footnotesize
% 	\generate{ delegate\_vec\_construction (V, $\alpha$)}{
% 				 \textit{//Phase I: warp-centric local maximum extraction}\\
% 				beg\_pos $\gets$ (warp\_id $<<$ $\alpha$) + lane\_id\;
% 				end\_pos $\gets$ ((warp\_id+1) $<<$ $\alpha$) + lane\_id\;
% 				% 	\While{beg\_pos < end\_pos}{
% 					\For{;\ beg\_pos < end\_pos;\ beg\_pos += 32}{
% 					max\_local $\gets$ \textbf{max} (max\_local, V[beg\_pos])\;
% 					}
					
% 			 \textit{//Phase II: intra-warp communication for subrange maximum}\\
% 			 \For{i = warp\_size>>1;\ i >=1;\ i = i >> 1}{
% 				max\_remote $\gets$ \textbf{shfl\_sync} (lane\_id+i)\;
% 				\uIf{lane\_id< i}{
% 					subrange\_max $\gets$ \textbf{max} (max\_local, max\_remote)\;			
% 				}
			
% 			}
			
% 			\textit{//Phase III: Save both subrange max and subrange id}\\ 
% 			\uIf{lane\_id==0}{
% 			subrange\_max $\gets$ subrange\_max\;
% 			subrange\_id $\gets$ warp\_id\;	
% 			%}
% %			$myWarpID \gets myWarpID+NumberOfWarps$\;
		
% 		    }
		
% 	}
% 	}
	
% % 	\generate{SOK()}{
% % 	$SampledArray \gets MaxSampling(OriginalArray,\alpha)$\;
% % 	$ContributingSubranges \gets RadixSelect(SampledArray)$\;
% % %	$NewArray \gets ConcatenateContributingSubranges(ContributingSubrangesArray)$\;
% % 	$NewArray \gets Concatenate(ContributingSubranges)$\;
% % 	$TopK \gets RadixSelect(NewArray)$\;	
% % }
% \end{algorithm}

\textbf{Warp-centric delegate vector construction} 
first divides the entire input vector into smaller subranges at the length of $2^\alpha$, where $\alpha$ is an integer. Afterward, each warp of GPU threads is assigned to extract each subrange delegate in three phases. Using maximum delegate as an example, every thread first records the maximum element when scanning through a specific subrange. Second, all the threads in each warp use \textit{shuffle instruction}, i.e., \_\_shfl\_sync(), to communicate and derive the maximum element in the current subrange. During the third phase, {\topk} writes each subrange's maximums and the subrange IDs to the delegate vector in the global memory. The size of the delegate vector is $\frac{|V|}{2^\alpha}$.

\textbf{Warp-centric concatenation.}
This step concatenates the eligible subranges into a new concatenated vector where the second top-$k$ performs on. Particularly, a warp of threads is responsible for moving the subrange elements into the concatenated vector. Because {\topk} uses delegate top-$k$ enabled filtering, the eligible elements per subrange are unclear. We resort to atomic operations to calculate the location for each eligible element. 
% \textbf{Second top-$k$} runs on the new vector that is constructed by the concatenation step and extracts the final top-$k$. Like first top-$k$, the second top-$k$ algorithm can be any of the popular top-$k$ algorithms. Different from the first top-$k$, this algorithm can be either top-$k$ or $k$-selection query. \fixme{remove first and second top-$k$?}

\textbf{First and second top-$k$.} Once {\topk} formulates delegate or concatenated vector, it will perform top-$k$ on them. While both top-$k$ algorithms work on a relatively small vector, the first top-$k$ presents two unique features. First, this top-$k$ algorithm has to work on a delegate vector that comes in the format of \textit{(key, value)} pair. Here, the \textit{key} is the delegate element from each subrange, and the \textit{value} indicates which subrange this delegate element comes from, which is essential for the concatenation step. Second, the first top-$k$ algorithm has to be a top-$k$ operation instead of $k$-selection because one needs to extract all the top-$k$ subranges for concatenation. We hence have to revise the radix and bucket $k$-selection algorithms of~\cite{alabi2012fast} to support top-$k$.

\textbf{Choice of top-$k$ algorithms.}
Despite the fact that {\topk} can help all existing top-$k$ algorithms, we notice that the best {\topk} will favor different top-$k$ algorithms when $k$ changes. Particularly, (i) when $k$ is small, all top-$k$ algorithms will enjoy comparable performance gains over their baseline algorithms. However, for radix and bucket top-$k$, they prefer in-place designs that always work on the input vector \textit{V} as instead of out-place variants that copy the derived candidates to a new array for the follow-up iteration. 
(ii) When $k$ is large, the performance of {\topk} assisted bitonic top-$k$ will lag behind. Specifically, bitonic top-$k$ needs a large shared memory space to cache the intermediate results and achieve desirable performance, which will experience low occupancy hence poor performance when $k>256$. 
\blue{As shown in Figure~\ref{fig:Inconsistency}, when $k$ goes beyond 256, the performance of bitonic top-$k$ degrades significantly. This makes {\topk} assisted bitonic perform worse than other {\topk} assisted ones.}

% Second, bucket top-$k$ cannot support optimized in-place top-$k$ which is viable in radix in-place select. It is important to mention that because both delegate and concatenated vectors are significantly smaller than the original vector $V$, a fast in-place top-$k$ algorithm is desirable.

\begin{figure}[t]
	\centering
\includegraphics[width=0.47\textwidth]{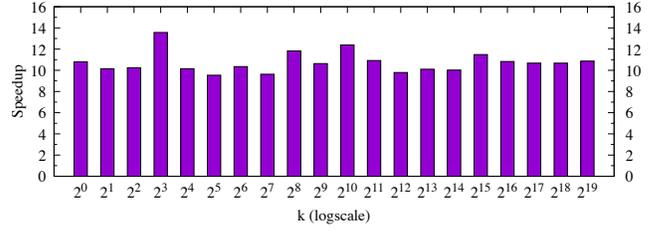}
% \vspace{-.15in}
	\caption{{\topk} in-place radix top-$k$ speedup over GGKS in-place radix top-$k$ on uniformly distributed vector of size $|V|=2^{21}$.
% 	\vspace{-.3in}
	}
	\label{fig:RadixOpt} 
\end{figure}

\textbf{Optimized in-place radix top-$k$}. 
% \fixme{Note: in-place bucket top-$k$ of GGKS doesn't modify the values} 
{Since existing in-place radix top-$k$ algorithm}~\cite{alabi2012fast} { requires to modify the ineligible element from the input vector into a value that is assured to fall out of the value range of interest (e.g., zero), this results in excessive random memory accesses. \textit{We introduce a single flag variable to indicate the radixes of interest.} This flag tracks the radixes that are eligible for the next iteration. 
Subsequently, once an element is loaded from global memory, we will perform flag} == (flag\ \&\ loadedElement) {between the loaded element and the flag variable.}  
{Only when the condition is evaluated as true, we consider this loaded element as a qualified element.} 
%%Commented on Apr 5
% {Although our current design focuses on integer values, this design can also be applied to other data types. 
%  After we find the qualified radixes in the current iteration, we will update the flag variable again to reflect our selections. 
% Updating the flag instead of $V$ avoids random global memory writes.}
%%Commented on Apr 5
{As shown in Figure}~\ref{fig:RadixOpt}, {our optimized 
in-place radix top-$k$ is on average 10.7$\times$ faster than the state-of-the-art}~\cite{alabi2012fast}.

% \fixme{We have also investigated an optimization for fast in-place bucket top-$k$, that is, using a value range, i.e., maximum and minimum, to indicate the elements of interest. Despite this optimization can reduce the time consumption for original in-place bucket top-$k$ (\fixme{how much}), this optimized in-place bucket top-$k$ introduces two logic operations as opposed to one in radix top-$k$. As a result, our optimized radix top-$k$ is an optimal choice for the first and second top-$k$.\update{Note: Our in-place bucket top-$k$ doesn't set a key to 0. It rather updates the max and min value based upon the updated bucket boundaries. At every iteration of reduction in in-place all the elements in the in-place method are checked if they fall within that max and min range. There is no modification of the values (like in-place radix top-$k$) in GGKS bucket top-$k$.}} Our evaluation shows that our in-place radix top-$k$ can reduce the time consumption for first and second top-$k$ from 34 ms to 28 ms when ??\fixme{}.
% the performance of radix top-$k$ in $U(0,2^{32})$ unsigned ints improved from approximately 34 ms to 28 ms.

% \vspace{-.1in}
\subsection{\texorpdfstring{$\alpha$}{} Tuning}\label{sec:alphaTune}

% A proper subrange size, i.e., $2^\alpha$, is crucial for {\topk} to achieve good performance: on one hand, if there are too many subranges, the delegate vector construction and the first top-$k$ would suffer from too many workloads. On the other hand, if there are too few subranges, majority of these subranges would be eligible for the second round. We hence skip too few subranges, leading to limited workload reduction for concatenation and the second top-$k$.

%Note use of semicolon indicates your intent to list items, however you only present one option. The expression "on one hand" is an introductory statement and blongs at the beginning of the sentence, perhaps: 
%\update{A proper subrange size is crucial for {\topk} to achieve good performance. On one hand, a small subrange may result in too many subranges introducing heavy workload during delegate vector construction and the first top-$k$.} 

A proper subrange size is crucial for {\topk} to achieve good performance. On the one hand, a small subrange size would lead to too many subranges. In this context, the delegate vector construction and the first top-$k$ would suffer from heavy workloads. On the other hand, when the subrange size is large, there are too few subranges. In this case, \update{the} majority of these subranges will be eligible for the second top-$k$. We hence skip too few subranges, leading to limited workload reduction for concatenation and second top-$k$. Rule~\ref{thm:alpha_tuning} helps {\topk} derive an optimal subrange size. 

\begin{myrule}\label{thm:alpha_tuning}
% 	To find the top-$k$ from a sequence of $|V|$ elements, 
% 	The $\alpha$ that yields optimal performance for {\topk} is 
For {\topk}, 
 	\begin{center}
$\alpha = \frac{1}{2} \cdot  [Const + \log_2({|V|}) -  \log_2({k})]$
 	\end{center} 
leads to the optimal subrange size $2^{\alpha}$, 
where $|V|$ is the number of elements in the input vector $V$, $k$ is the number of top elements {\topk} aims to find. 
% $Const =  \log_2[5\cdot( C_{global}+C_{shfl})]-  \log_2(6\cdot C_{global})$, 
$Const =  \log_2[6\cdot( C_{global}+C_{shfl})]-  \log_2(6\cdot C_{global})$, 
where $C_{global}$ and $C_{shfl}$ are the clock cycles for one global memory access and one CUDA shuffle instruction, respectively. 
% Note, we assume the subrange size is $2^{\alpha}$.
\end{myrule}

% \fixme{A reviewer feedback: $C_{global}$ and $C_{shfl}$ do not take into account the parallelism of the GPU. Cglobal usage does not take into account that warps are executed in parallel, so the result can conceivably change with number of SMs. Cshfl usage does not take into account that entire warp is executed in lock-step. The difference is moved to the constant, and this calculation is rather vague.}

% \fixme{The optimum value of alpha depends on Cglobal and Cshfl that may change with DDR/GPU improvements. So the value might need to be retuned for other GPU families. There is no analysis to verify if the optimum calculation/value remains constant over GPUs, or if there is a range that is valid for current GPU families.}

\begin{proof} 
The time consumption of {\topk} is: 
\begin{equation} \label{eq:T_total}
\small
\setlength{\belowdisplayskip}{5pt} \setlength{\belowdisplayshortskip}{5pt}
\setlength{\abovedisplayskip}{5pt} \setlength{\abovedisplayshortskip}{5pt}
T  ={T_{Delegate} + T_{FirstK} + T_{Concat} + T_{SecondK}},
\end{equation}
where $T_{Delegate}$, $T_{FirstK}$, $T_{Concat}$ and $T_{SecondK}$ are the time consumption of delegate vector construction, first top-$k$, concatenation, and second top-$k$, respectively.

Global memory access and intra-warp communication are the key factors determining the time consumption of {\topk} for two reasons. First, one global memory access or intra-warp shuffle operation takes a much longer time than a single arithmetic and logic operation on GPUs, according to Nvidia profiler~\cite{nvprof}. Second, the number of arithmetic and logical operations is similar to that of global memory accesses across all four stages of {\topk}. Using delegate vector construction as an example, each thread loads one element from global memory and compares, i.e., a logic operation, it with the current maximum in a register. In this case, one memory access leads to one arithmetic or logic operation. As a result, we mainly use global memory access and shuffle instructions to estimate the time consumption. We perform our analysis for maximum delegate for simplicity and assume all global memory accesses have equal latency, i.e., $C_{global}$.

% \fixme{Better use Mem instead of IO, IO is not about memory access, it refers to other devices. Better use throughput instead of C, it's what actually decides the GPU performance because latencies are overlapped with each other heavily. Same for shl. Also mention you are assuming equal cost for read and write.}

\textit{$T_{Delegate}$:}
Delegate vector construction reads $|V|$ elements and write $\frac{|V|}{2^{\alpha}}$ delegates. After thread local comparison, each subrange resorts to CUDA \_\_shfl\_sync instruction to derive the maximum for the entire subrange. Since one warp contains 32 threads, $\sum_{1\le i\le 5}\frac{32}{2^i} = 31$ shuffle instructions are needed. Therefore, the communication complexity is $31\cdot\frac{|V|}{2^{\alpha}}\cdot C_{shfl}$, where $C_{shfl}$ is the cost of a shuffle instruction. 
Together, delegate vector construction time is: %\\
\begin{equation}\label{eq:T_sample}
\small
\setlength{\belowdisplayskip}{5pt} \setlength{\belowdisplayshortskip}{5pt}
\setlength{\abovedisplayskip}{5pt} \setlength{\abovedisplayshortskip}{5pt}
T_{Delegate} =\underbrace{(1+\frac{1}{2^{\alpha}})\cdot |V|\cdot C_{global}}_{Global\ memory\ access} +\underbrace{\frac{31\cdot |V|}{2^{\alpha}}\cdot C_{shfl}}_{Intra-warp\ comm.}.
%or, T_{sampling} = |V|\cdot [C+(C+5\cdot C_{IO})\cdot 2^{-\alpha}]
\end{equation}

% $T_{FirstK}$: Now, we analyze the time consumption of our optimized in-place radix top-$k$. According to our study, eight bits per digit yields the optimal performance for in-place optimized radix top-$k$. A 32-bit unsigned integer hence experiences four iterations of scans. {At each iteration, we always load all the elements. In the end, we write $k$ elements, which are the indices of the eligible subranges.} Therefore, the time consumption of the first top-$k$ is: 
$T_{FirstK}$: Now, we analyze the time consumption of our optimized in-place radix top-$k$. According to our study, 8-bit per digit yields the optimal performance for in-place optimized radix top-$k$. A 32-bit unsigned integer hence experiences four iterations of scans. {At each iteration, we always load all the elements. An additional iteration over the vector is used to identify the top-k elements. Finally, we write $k$ elements, which are also the indices of the eligible subranges.} Therefore, the time consumption of the first top-$k$ is: 
% \fixme{key value pair?}
\begin{equation}\label{eq:T_firstTopK}
\small
\setlength{\belowdisplayskip}{5pt} \setlength{\belowdisplayshortskip}{5pt}
\setlength{\abovedisplayskip}{5pt} \setlength{\abovedisplayshortskip}{5pt}
% T_{FirstK}= \frac{4\cdot  |V|\cdot C_{global}}{2^{\alpha}} + k\cdot C_{global}.
T_{FirstK}= \frac{5\cdot  |V|\cdot C_{global}}{2^{\alpha}} + 2\cdot k\cdot C_{global}.
\end{equation}

\begin{figure}[t]
 	\centering
 	\includegraphics[width=.47\textwidth]{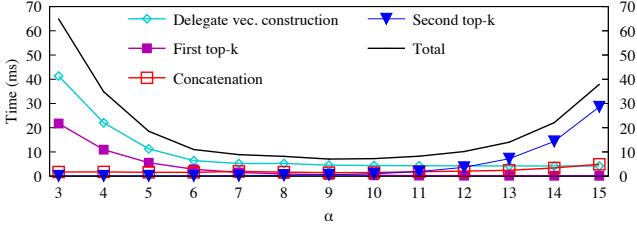} 
%  	\vspace{-.1in}
 	\caption{The runtime of {\topk} with respect to the change of $\alpha$, where $k=2^{13}$ and the UD dataset from Section~\ref{sec:evaluation}. 
%  	\vspace{-.1in}
 	}
 	\label{fig:SOKTimeAsFunctionofAlpha} 
 \end{figure}

$T_{Concat}$: The concatenation step reads $k$ indices for the subranges that are eligible for the second top-$k$ and copies those subranges from the input to the concatenated vector for the second top-$k$. The time consumption for concatenation is:

\begin{equation}\label{eq:T_concatenation}
\small
\setlength{\belowdisplayskip}{5pt} \setlength{\belowdisplayshortskip}{5pt}
\setlength{\abovedisplayskip}{-5pt} \setlength{\abovedisplayshortskip}{-5pt}
T_{Concat} = k\cdot C_{global}+ 2\cdot k\cdot 2^{\alpha}\cdot C_{global}.
\end{equation}
%$\\So, T_{Concat}=\frac{5CK2^{\alpha + 5}}{N_T}\\$

$T_{SecondK}$: The second top-$k$ takes as input the output from the concatenation step and conducts in-place radix top-$k$ to derive the eventual top-$k$. Consequently, this step is mainly about reading the entire outputs from concatenation. Similar to the analysis for the first top-$k$, which reads the concatenated vector by four times, the time consumption of the second top-$k$ is:
\begin{equation}\label{eq:T_SecondTopK}
\small
\setlength{\belowdisplayskip}{5pt} \setlength{\belowdisplayshortskip}{5pt}
\setlength{\abovedisplayskip}{5pt} \setlength{\abovedisplayshortskip}{5pt}
T_{SecondK}=4 \cdot k \cdot 2^{\alpha}\cdot C_{global}.
\end{equation}

Taken Equations~(\ref{eq:T_sample}) --~(\ref{eq:T_SecondTopK}) together, we arrive at the total time consumption of {\topk} as shown as in Equation~\ref{eq:Total}.

% \vspace{-.1in}
\begin{align}\label{eq:Total}
\begin{split}
\small
% \setlength{\belowdisplayskip}{5pt} \setlength{\belowdisplayshortskip}{5pt}
% \setlength{\abovedisplayskip}{-15pt} \setlength{\abovedisplayshortskip}{-15pt}
% T&={T_{Delegate} + T_{FirstK} + T_{Concat} + T_{SecondK}}\\
% &=31\cdot |V| \cdot 2^{-\alpha}\cdot  C_{shfl} +\\ 
% &(5\cdot|V|\cdot 2^{-\alpha}+6 \cdot k\cdot 2^{\alpha} +2 \cdot k +  |V|)\cdot C_{global}.
T&={T_{Delegate} + T_{FirstK} + T_{Concat} + T_{SecondK}}\\
&=31\cdot |V| \cdot 2^{-\alpha}\cdot  C_{shfl} +\\ 
&(6\cdot|V|\cdot 2^{-\alpha}+6 \cdot k\cdot 2^{\alpha} +2 \cdot k +  |V|)\cdot C_{global}.
\end{split}
\end{align}

Given Equation~\ref{eq:Total} ignores various hardware scheduling, arithmetic, and logical operation latency, we introduce $\Delta(\alpha, k, |V|)$ (at Equation~\ref{eq:Total_timeDefault}), which is a positive function of $\alpha$, $k$ and $|V|$, to make up the impacts. We assume the magnitude of $\Delta(\alpha, k, |V|)$ is smaller than that of $T$.

% \begin{equation}\label{eq:Total}
% \begin{split}
% \small
% \setlength{\belowdisplayskip}{5pt} \setlength{\belowdisplayshortskip}{5pt}
% \setlength{\abovedisplayskip}{0pt} \setlength{\abovedisplayshortskip}{0pt}
% T ={T_{Delegate} + T_{FirstK} + T_{Concat} + T_{SecondK}}\\
% &=31\cdot |V| \cdot 2^{-\alpha}\cdot  C_{shfl} +\\ 
% &(5\cdot|V|\cdot 2^{-\alpha}+6 \cdot k\cdot 2^{\alpha} +2 \cdot k +  |V|)\cdot C_{global}.
% \end{split}
% \end{equation}

% Given Equation~\ref{eq:Total} ignores various hardware scheduling, arithmetic, and logical operation latency, we introduce $\Delta(\alpha, k, |V|)$, which is a positive function of $\alpha$, $k$ and $|V|$, to make up the impacts. We assume the magnitude of $\Delta(\alpha, k, |V|)$ is smaller than that of $T$. 

\begin{equation}\label{eq:Total_timeDefault}
% \begin{split}
\small
\setlength{\belowdisplayskip}{5pt} \setlength{\belowdisplayshortskip}{5pt}
\setlength{\abovedisplayskip}{-5pt} \setlength{\abovedisplayshortskip}{-5pt}
T_{\topk} = T + \Delta(\alpha, k, |V|).
% &=(31\cdot  C_{shfl} + 5\cdot C_{global})\cdot |V| \cdot 2^{-\alpha}\\
% &+6 \cdot k\cdot C_{global}\cdot 2^{\alpha} +2 \cdot k\cdot C_{global} \\
% &+  |V|\cdot C_{global}+\Delta(\alpha, k, |V|).
% \end{split}
\end{equation}
% Below we analysis Equation~\ref{eq:Total_timeDefault} to extract the sweetspot of $\alpha$.
% Recalling the observation from Figure~\ref{fig:SOKTimeAsFunctionofAlpha}, the performance of {\topk} is a convex function of $\alpha$. Our analysis is also inspired by this observation. Particularly, 

\textit{We first prove $T_{\topk}$ is a convex function, which makes it easy to obtain the optimal $\alpha$ for {\topk}.}
% In order to achieve the optimal performance from {\topk}, we need to minimize the $T$ from the Equation~\ref{eq:Total_timeDefault}. 
According to~\cite{minima}, in order to demonstrate the convex nature of $T_{\topk}$, the second derivative of $T_{\topk}$ with respect to $\alpha$ should be positive. 
% The second order partial derivative of $T_{total}$ with respect to $\alpha$ is
\begin{align}\label{eq:2nd}
\small
\begin{split}
\small
\setlength{\belowdisplayskip}{5pt} \setlength{\belowdisplayshortskip}{5pt}
\setlength{\abovedisplayskip}{-5pt} \setlength{\abovedisplayshortskip}{-5pt}
% \frac{\partial^{2} T_{\topk}}{\partial^{2} \alpha}&= (31\cdot C_{shfl}+ 5\cdot C_{global} )\cdot  |V|\cdot \ln^2(2)\cdot 2^{-\alpha}\\
% &+6 \cdot k\cdot C_{global}\cdot \ln^2(2)\cdot 2^{\alpha} + \Delta''(\alpha, k, |V|).
\frac{\partial^{2} T_{\topk}}{\partial^{2} \alpha}&= (31\cdot C_{shfl}+ 6\cdot C_{global} )\cdot  |V|\cdot \ln^2(2)\cdot 2^{-\alpha}\\
&+6 \cdot k\cdot C_{global}\cdot \ln^2(2)\cdot 2^{\alpha} + \Delta''(\alpha, k, |V|).
\end{split}
\end{align}
%$\\\frac{\partial^{2} T_{total}}{\partial^{2} \alpha}= (31\cdot C_{shfl}+ 5\cdot C_{global} )\cdot  |V|\cdot (\log_{2}(2))^{2}\cdot 2^{-\alpha}+2 \cdot k\cdot C_{global}+6 \cdot k\cdot C_{global}\cdot (\log_{2}(2))^{2}\cdot 2^{\alpha}\\$
According to the assumption in Equation~\ref{eq:Total_timeDefault}, the magnitude of $\Delta''(\alpha, k, |V|)$ will be smaller than the remaining factors in Equation~\ref{eq:2nd}. For positive values of $k$, $|V|$, $C_{global}$ and $C_{shfl}$, we obtain: 
\begin{equation}
\small
    \frac{\partial^{2} T_{\topk}}{\partial^{2} \alpha} > 0.
\end{equation}
Hence, $T_{\topk}$ is convex function of $\alpha$.

% \textbf{Observation}. https://www.overleaf.com/project/5cec16ff6eeaa05db8c5c19c

Our study in {Figure~\ref{fig:SOKTimeAsFunctionofAlpha} also suggests that {\topk} is a {convex} function of $\alpha$}. Particularly, the time consumption of delegate vector construction and first top-$k$ decrease along with the increase of $\alpha$. Meanwhile, concatenation and second top-$k$ increase. Altogether, the total time consumption decreases then increases with respect to the increase of $\alpha$. Finally, since $T_{\topk}$ is convex, the optimal value of $\alpha$ can be obtained by:
% differentiating $T$ with respect to $\alpha$ and solving for $\alpha$ with first derivative as 0. 
% Refer to Figure~\ref{fig:SOKTimeAsFunctionofAlpha} which is demonstrating the minima of the curve of total time.
%\begin{center}
 \begin{figure}[t]
	\centering
	\includegraphics[width=.47\textwidth]{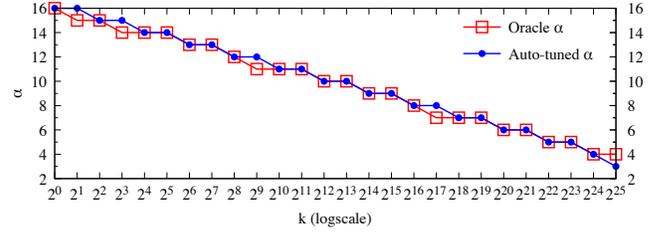}
% 	\vspace{-.1in} %1.png is the relative path of image
	\caption{ Performance of oracle $\alpha$ vs. auto-tuned $\alpha$.
% 	\vspace{-.1in}
	} %Number of elements=$2^{29}$  unsigned integers.} %caption is the title of image
	\label{fig:SOK-Comparison_ManualAlphatuneVsAdap_AlgAlpha_tune} %  used for citing
	%   \vspace{-.1in}
\end{figure}
\begin{equation}\label{eq:eq0}
\frac{\partial T_{\topk}}{\partial \alpha} = 0. 
% \frac{\partial T_{Delegate}}{\partial \alpha} + \frac{\partial T_{FirstK}}{\partial \alpha} + \frac{\partial T_{Concat}}{\partial \alpha} + \frac{\partial T_{SecondK}}{\partial \alpha}.
\end{equation}%\end{center}
Solving Equation~\ref{eq:eq0}, we obtain: 
% For maxima or minima, 
% $\frac{\partial T}{\partial \alpha} =0 $, solving for $\alpha$
\begin{center}
	\begin{equation}\label{eq:tunedalpha}
	\small
\setlength{\belowdisplayskip}{5pt} \setlength{\belowdisplayshortskip}{5pt}
\setlength{\abovedisplayskip}{-10pt} \setlength{\abovedisplayshortskip}{-10pt}
	\boxed{\alpha = \frac{1}{2} \cdot  [\log_{2}({|V|}) - \log_{2}({k}) + const]},
	\end{equation}
\end{center}	
% where, $const = \log_{2}(5\cdot C_{global}+31\cdot C_{shfl})- \log_{2}(6\cdot C_{global}) +\Delta'(\alpha, k, |V|)$.
where, $const = \log_{2}(6\cdot C_{global}+31\cdot C_{shfl})- \log_{2}(6\cdot C_{global}) +\Delta'(\alpha, k, |V|)$.
%for unsigned int let, $C_{IO}=4\cdot C$.
% \begin{center}\label{eq:tunedalpha}
% 	
% 	\begin{equation}
% \therefore	\boxed{\alpha = \frac{1}{2} \cdot  [\ln({|V|}) - \ln({k})-2.1]}
% 	\end{equation}
% \end{center}	
% In order to extract a practical $\alpha$, we further derive the correct $const$ value by testing {\topk} across various $k$ and data distributions. 
% Figure~\ref{fig:const_tuning} suggests that const = 2 is the optimal option across a wide range of $|V|$ and $k$.

% \begin{figure}[h]
% 	\centering
% 		\vspace{-.2in}
% 	\begin{tabular}{cc}
% 		\hspace{-.1in}\subfloat[{$k=2^{25}$}]{\includegraphics[width=0.24\textwidth]{./figures/const_tuning.pdf}} & 
% 		\hspace{-.1in}\subfloat[{$|V|=2^{30}$}]{\includegraphics[width=0.24\textwidth]{./figures/const_tuningK.pdf}}\\
% 	\end{tabular}
% 	\vspace{-.15in}
% % 	\caption{The performance dynamics with respect to the change of constant value in Equation~\ref{eq:tunedalpha} when (a) varying $V$ at $|k|=2^{20}$, and (b) varying $k$ at $V=2^{30}$. 
% 	\caption{The performance dynamics with respect to  \textit{const} in Equation~\ref{eq:tunedalpha} when (a) varying $V$ at $|k|=2^{20}$, and (b) varying $k$ at $V=2^{30}$.
% \vspace{-.3in}
% 	}
% 	\label{fig:const_tuning}
% % 	\vspace{\shrink}
% \end{figure}

% Figure~\ref{fig:SOK-Comparison_ManualAlphatuneVsAdap_AlgAlpha_tune}, from an evaluation perspective, exhibits the performance alignment of the auto-tuned $\alpha$ and the oracle $\alpha$ across a wide range of $k$ for the $|V|=2^{29}$ unsigned integers dataset, where we set const = 3 according to performance tuning.
Figure~\ref{fig:SOK-Comparison_ManualAlphatuneVsAdap_AlgAlpha_tune}, from an evaluation perspective, exhibits the performance alignment of the auto-tuned $\alpha$ and the oracle $\alpha$ across a wide range of $k$ for the $|V|=2^{30}$ unsigned integers dataset, where we set const = 3 according to performance tuning.
% , identical to that of Figure~\ref{fig:SOKTimeAsFunctionofAlpha}.
\end{proof}

% \vspace{-.1in}
\subsection{Delegate Vector Construction Optimization}\label{sub:sample_optimization}

After we optimize the first and second top-$k$ computations and concatenation steps, delegate vector construction becomes the next bottleneck for {\topk}. 
This is especially true when $k$ is relatively large. According to Equations~\ref{eq:tunedalpha}: $\alpha$ decreases with respect to the increase of $k$. For instance, when |V|=$2^{30}$, and $k=2^{24}$, the optimal $\alpha = 4$. This implies that the input vector is partitioned into a large number of small subranges which would lead to two problems: \textit{(i) the small subrange size fails to saturate a GPU warp; and (ii) too many subranges will lead to an overwhelming number of shuffle instructions for delegate communication.}

We introduce a novel \textbf{coalesced loading to shared memory and strided computing approach} to remedy this small subrange size problem ($\alpha \leq 5$). This method consists of two phases: (i) one warp moves 32 subranges into the shared memory for delegate extraction. Here, each subrange is loaded from global memory into the shared memory by a warp in a coalesced manner. Since the subrange size is small, the shared memory pressure remains low. Subsequently, (ii) each thread of the warp individually works on the entire subrange to extract the delegate. This design ensures that all the threads of a warp have workloads, and no shuffle instruction is needed to communicate and decide the subrange delegates. This design helps the $\beta$ delegate tremendously, which would otherwise needs approximately $\beta\times$ more shuffle instructions to extract the $\beta$ delegates. We use padding to avoid shared memory bank conflict.

\begin{figure}[t]
	\centering
\includegraphics[width=0.47\textwidth]{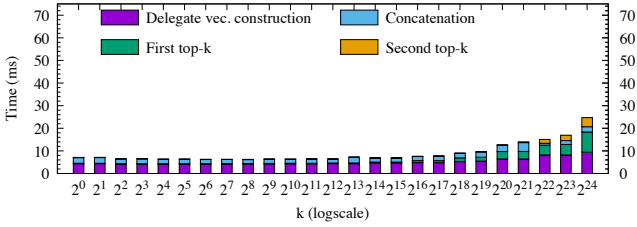}
% \vspace{-.15in}
% 	\caption{After delegate vector construction optimization, {\topk} time consumption breakdown for the $2^{30}$ elements that are uniformly generated, which is identical to the dataset in Figure~\ref{fig:SOK_decomposition_default}.
	\caption{After delegate vector construction optimization, {\topk} time consumption breakdown for UD dataset on Section~\ref{sec:evaluation}.
% 	\fixme{consistent}
% 	\vspace{-.2in}
	}
	\label{fig:SOK_decomposition_delegate_opt} 
\end{figure}

Figure~\ref{fig:SOK_decomposition_delegate_opt} shows the improvement brought by the delegate vector construction optimization for different values of $k$. Comparing to Figure~\ref{fig:opt_concat_Uniform_With_Beta_split}, one can find out that the delegate vector construction time is dramatically reduced for larger values of $k$, making the sampling time always close to merely the time consumption of scanning the input vector. Especially, for $k=2^{24}$, we observe the time consumption of delegate vector construction decreases from \update{31.4} ms to \update{9.4} ms. And the total time consumption is reduced from 46.7 ms of Figure~\ref{fig:opt_concat_Uniform_With_Beta_split} to 24.7 ms.
 
%  The improvement in mainly seen in the small values of $\alpha$ which is expected as for large values of $\alpha$, it is already near to the scanning bandwidth of approximately 4 ms. The shared memory implementation in maximum sampling help to push the  {\topk} to be better than the normal radix top-$k$ for higher values of k.
% The  {\topk} performance with addition of the sampling optimization for small subranges ($\alpha \leq 5$) is demonstrated in Figure~\ref{fig:SOK_decomposition}. The algorithm for optimized sampling kernel is shown Algorithm~\ref{alg:Max_sharedMem}.

\subsection{Distributed {\topk}}\label{sub:multigpu}

In the distributed GPU setting, we partition the input vector $V$ into disjoint sub-vectors of equal length. To fit in the GPU memory, we require the length of each sub-vector to be no longer than $2^{30}$: (i) when $(\#GPUs)\times 2^{30}\ge |V|$, we partition V into \#GPU number of sub-vectors and let each GPU account for one sub-vector. (ii) When $(\#GPUs)\times 2^{30}< |V|$, we partition $V$ into $\frac{|V|}{2^{30}}$ number of sub-vectors. In this case, one GPU accounts for more than one sub-vector; hence will load the unloaded sub-vectors from outside of GPUs. We schedule each GPU to compute the top-$k$ for its own sub-vectors to arrive at one top-$k$ per GPU. Subsequently, each GPU sends its top-$k$ to the primary GPU to calculate the final top-$k$.

\begin{wrapfigure}{r}{0.25\textwidth}
% \vspace{-20pt}
% \hspace{50pt}
  \begin{center}
  	\includegraphics[width=0.25\textwidth]{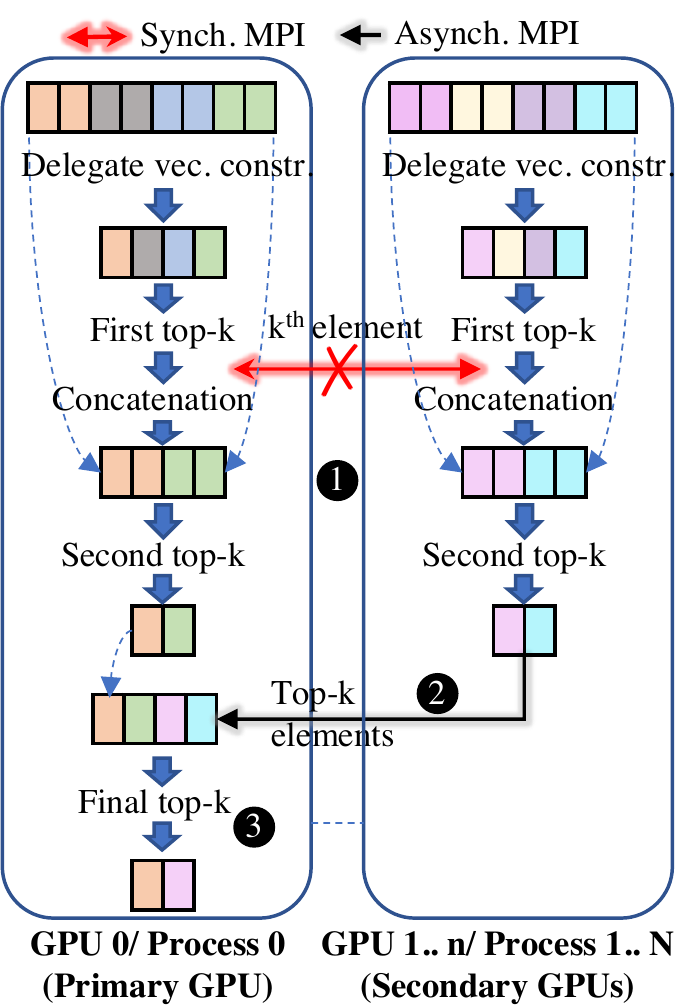}
  \end{center}
%   \vspace{-12pt}
	\caption{\blue{Workflow of multi-GPU {\topk}.}}
% 	We assume the input vectors for top-$k$ computation are in the GPUs}.}
% 	\vspace{-12pt}
	\label{fig:multiGpu_arch} 
\end{wrapfigure}

\blue{Figure~\ref{fig:multiGpu_arch} presents the workflow of multi-GPU {\topk} which contains three major steps: \circled{1} It enables all the participating GPUs to work on their local sub-vectors to compute local top-$k$. \circled{2} It gathers these locally computed top-$k$'s to the primary GPU. \circled{3} It enables primary GPU to compute the global top-$k$. For inter-GPU communication, we use Message Passing Interface (MPI)~\cite{snir1998mpi}. Particularly, we use MPI asynchronous (\textcolor{black}{$\leftarrow$} \textit{Asynch. MPI} in the figure) communication among the processes to gather local top-$k$'s from all the GPUs to the primary GPU. 
}

While relying on the primary GPU to compute the final top-$k$ works for a small number of GPUs, e.g., 16 in our evaluation, we anticipate hierarchical reduction~\cite{yangtree} would excel when {\topk} scales to a large number of GPUs. 
% The second technique uses blocking binary tree reduction~\cite{yangtree} method for gathering top-$k$ elements from each GPUs to primary GPU after second top-$k$. We explore the method for reducing the network congestion during communication of top-$k$.
Particularly, for a multi-node setting, where each node installs multiple GPUs, the hierarchical scheduling method first derives the top-$k$ across all the GPUs in each node.
Afterward, all the nodes will send their top-$k$ to the primary GPU to compute the final top-$k$.

% the communication first involves gathering of top-$k$s' among intra-node GPUs to a GPU of the node. The GPU that has the top-$k$s' of all intra-node GPUs then participates in reducing the top-k to the primary GPU in primary node. The primary GPU then computes the final top-k. 
%   Such reduction of top-$k$ follows binary tree reduction.
%   As such communication requires the necessity of synchronization among the GPUs the benefit for reducing the network congestion should exceed the overhead of synchronization. For small $k$, as the communication volume remains low {\topk} performs better when it uses asynchronous communication of top-$k$. However, for large $k$ as the network congestion increases this method of reducing the network congestion can become beneficial.} 
  
% Third, during computation of top-$k$ local to a GPU, {\topk} considers a case when the GPU memory is not enough for storing all the input elements locally ($|V| > 2^{30}$). 
% In this context, we will partition 

% In such case, {\topk} considers the excessive input elements are in CPU memory. After the GPU finishes processing the input elements in its memory, {\topk} reloads the elements present in CPU to GPU memory. Note, the overhead of reloading the elements can become significant. We can observe such overheads in Table~\ref{tab:multiGPU} at Section~\ref{sec:evaluation}.} 

\blue{
%We further investigate two techniques to improve the performance of multi-GPU version of {\topk}. 
It is also worthy of noting that we attempt to reduce the workload for the second top-$k$ by enabling an MPI communication for the top-$k^{th}$ element of the first top-$k$ (the \textcolor{red}{$\leftrightarrow$} symbol in Figure~\ref{fig:multiGpu_arch}). With the $k^{th}$ delegate across all GPUs, we anticipate this will help filter out more unpromising elements hence reduce the workload.
% MPI in the Figure) among GPUs. This technique takes the maximum of $k^{th}$ elements from all the GPUs and use it as an updated $k^{th}$ element during the concatenation phase. On looking back at Rule~\ref{rule:filter}, concatenation uses $k^{th}$ element of first top-$k$ for the \textit{delegate top-$k$ enabled filtering optimization} to only consider the elements that are larger than the $k^{th}$ element for the second top-$k$.  
% As per Rule~\ref{rule:filter}, 
% As the GPU that contributes to the maximum $k^{th}$ element in this technique already takes at least $k-1$ elements greater than the $k^{th}$ element for second top-$k$, we can decide that the elements smaller than this maximum $k^{th}$ element in other GPUs will not contribute to final top-$k$. This ensures the correctness of the technique. The technique helps concatenation to further reduce concatenated vector size, which as a result leads to  performance improvement of second top-k. 
However, since this method requires all GPUs to have the maximum of $k^{th}$ elements before launching the concatenation kernel, this introduces synchronization overhead. Additionally, in Figure~\ref{fig:SOK_decomposition_delegate_opt}, we also notice the cost of second top-$k$ remains low throughout for a wide range of $k$, leaving relatively small room for improvement. 
In summary, the overhead of synchronizing the $k^{th}$ elements from first top-$k$s' exceeds the benefits of a smaller concatenated vector, we disable this technique in our final version of distributed {\topk}.
} 

\section{Evaluation}
% \td Suggestion for using a real dataset for comparison \td.
\label{sec:evaluation}

We implement {\topk} with $\sim$1,500 lines of C++ and CUDA code, extending the state-of-the-art bitonic, bucket and radix top-$k$ projects~\cite{shanbhag2018efficient,alabi2012fast}. We compile the source code using NVIDIA CUDA 10.1 nvcc compiler with the optimization level as O3.
{We use two platforms to evaluate the performance of {\topk}.
Platform I is a server with two Intel Xeon ``Cascade Lake-SP'' CPUs (@3.8 GHz) and 4 Tesla V100S GPU running Ubuntu Server 18.04. 
Platform II consists of i7-8700 CPU @ 3.20GHz with one Titan Xp running Ubuntu Server 16.04. 
} All the reported execution time is an average of five runs. The default size of the input vector $V$ is $|V|=2^{30}$, and each data entry is an unsigned integer. $V$ is generated by the following distributions. 

\begin{itemize}
    \item \textbf{Uniform distribution dataset (UD)} is generated {following U[0, $2^{32}$-1]}, meaning the value ranges from 0 to $2^{32}-1$.
    \item \textbf{Normal distribution dataset (ND)} is {produced with the normal distribution $N[10^8, 10]$}, where the mean and standard deviation are $10^8$ and 10, respectively.
      \item \blue{\textbf{Customized distribution dataset (CD)} is produced to increase the number of iterations in bucket top-$k$. The values are generated in the range of [0, $2^{32}$-1] such that every bucket other than the bucket containing the $k^{th}$ element will always have at least one element in every iteration and majority of the element is present in the bucket with the $k^{th}$ element.}
    %   Once the values no longer be written for single element per bucket for an iteration, the remaining vector part is filled with  $2^{32}$-1.}
\end{itemize} 
Unless stated differently, we present the experiments on platform I for the UD dataset.

\subsection{{\topk} vs. State-of-the-art}\label{sub:SOAVsSOK}

This section reports the performance gains brought by {\topk} to the state-of-the-art with respect to the change of $|V|$ and $k$.  

\begin{figure}[t]
	\centering
	\includegraphics[width=.47\textwidth]{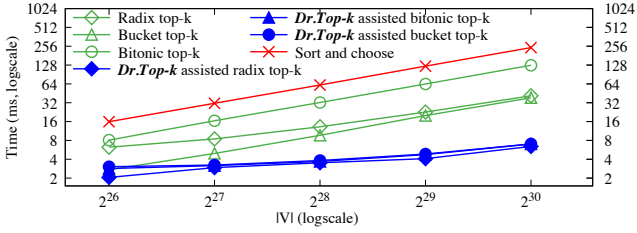} 
% 	\vspace{-.1in}%1.png is the relative path of image
	\caption{\blue{The time consumption of {\topk} versus various top-$k$ algorithms with respect to the increase of $|V|$.}
% 	\fixme{We need {\topk} to be always faster \update{The curves are better in 128 than in 256. Probably we can set new |V| range from $2^{27} to 2^{30}$? Note: The test for $|V|=2^{31}$ failed for both GGKS and SIGNOD'18 bitonic top-k.}}
% 	\fixme{Key to be {\topk} assisted?}
% 	\fixme{All the keys should be top-$k$ as opposed to $k$-selection?}
% 	\vspace{-.1in}
	} 
	\label{fig:DiffN_K_128Comparison} %  used for citing
% 	  \vspace{-.1in}	
\end{figure}

%Commented on May 24
% \textbf{{\topk} for different input vector $V$ sizes.} Figure~\ref{fig:DiffN_K_128Comparison} demonstrates the time consumption of {\topk} for $k=1024$ on $V$ whose sizes vary from {$2^{26}$} to $2^{30}$. \textit{The general trend is that {\topk} becomes more beneficial when the input vector size is bigger}, because delegate vectors can help reduce more workloads when $V$ gets larger. Particularly, when $|V|=2^{30}$, radix, bucket and bitonic top-$k$ consume {41.3 ms, 38.4 ms and 127.0 ms}, respectively. Our {\topk} assisted radix, bucket and bitonic top-$k$ designs reduce the time to {6.4 ms, \update{7.0} ms and 7.0 ms}, respectively. 
%Commented on May 24

\textbf{{\topk} for different input vector $V$ sizes.} Figure~\ref{fig:DiffN_K_128Comparison} demonstrates the time consumption of {different versions of top-$k$} for $k=1024$ on $V$ whose sizes vary from {$2^{26}$} to $2^{30}$. \textit{The general trend is that {\topk} becomes more beneficial when the input vector size is bigger}, because delegate vectors can help reduce more workloads when $V$ gets larger. Particularly, when $|V|=2^{30}$, radix, bucket, bitonic \blue{and sort and choose top-$k$} consume {41.3 ms, 38.4 ms, 127.0 ms \blue{and 243.2 ms}}, respectively. Our {\topk} assisted radix, bucket and bitonic top-$k$ designs reduce the time to {6.4 ms, \update{7.0} ms and 7.0 ms}, respectively.

\begin{figure}[t]
	\centering
		\includegraphics[width=.47\textwidth]{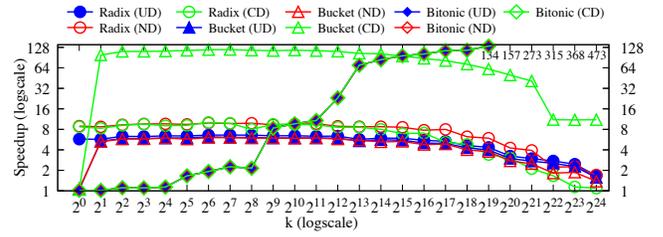}
% 	\includegraphics[width=.47\textwidth]{./figures/all_speedup_broken_des.pdf}
% 	\vspace{-.15in} 
	\caption{The speedup of {\topk} over the-state-of-the-art for a varying $k$ on synthetic datasets.
% 	\vspace{-.1in}
	} 
	\label{fig:all_speedup} 
% 	  \vspace{-.1in}	
\end{figure}

\textbf{{\topk} assisted radix top-$k$.} As shown in Figure~\ref{fig:all_speedup}, in general, {\topk} yields bigger performance gains on the  normal {and customized distribution} datasets. Particularly, we observe $1.7\times$ - $10\times$ and {$1.1\times$ - $10.1\times$} speedups, {respectively}  on normal and customized distribution, while \update{$1.7\times$ - $6.6\times$} on uniform distribution. It is also important to note that the impact of {\topk} decreases with respect to the increase of $k$. For instance, when k=$2^{24}$, {\topk} only gives $1.7\times$ speedups for both the uniform and normal distribution datasets {and 1.1$\times$ speedup for customized distribution}. This is caused by the fact that {\topk} requires more delegates to figure out the useful subranges when $k$ becomes larger, leading the first top-$k$ to consume significant time. Section~\ref{subsec:stat} conducts a thorough study on the workload reduction trend for varying $k$.

\textbf{{\topk} assisted bucket top-$k$.} Figure~\ref{fig:all_speedup} also shows the speedup of {\topk} assisted bucket top-$k$ over the bucket top-$k$ alone algorithm on the normal, {customized} and uniform distributions. The trends are analogous to radix top-$k$ but with two differences. First, bucket top-$k$ performs fairly well when $k=1$ because bucket top-$k$ first finds the maximum value. Then the query is completed. 
% Thanks to $\beta$ delegate, {\topk} assisted bucket top-$k$ only needs to work on the first top-$k$, leading {\topk} assisted bucket top-$k$ to perform similar, i.e., by $1.1\times$ for all normal, {customized} and uniform distributions.
Thanks to near bandwidth performance of delegate vector construction and single delegate needed to be selected for first top-$k$ for $k=1$, {\topk} assisted bucket top-$k$ only needs to work on the first top-$k$, leading {\topk} assisted bucket top-$k$ to perform faster, i.e., by $1.1\times$ for all normal, {customized} and uniform distributions.
 Second, in bucket top-$k$, the speedup of {\topk} on {customized} distribution outpaces the uniform and normal distribution. 
 Particularly, {we observe the speedups from 
 $1.1 \times$ - $118.6 \times$ for customized distribution}
 while $1.1 \times$ - $6.1 \times$ and $1.1 \times$ - \update{$6.2 \times$} for normal and uniform distribution respectively. 

\textbf{{\topk} assisted bitonic top-$k$.} Figure~\ref{fig:all_speedup} further includes the speedup of {\topk} assisted bitonic top-$k$ over the bitonic top-$k$ stand alone algorithm~\cite{shanbhag2018efficient}. {Note, the original source code}~\cite{bitonic_select_src} {from bitonic top-\textit{k} project}~\cite{shanbhag2018efficient} {experiences shared memory overflow when $k$ goes beyond 256. We modify the source code to enable it for $k>256$.} Particularly, the speedup of {\topk} climbs from \update{$1.1 \times$ when k=$2^{0}$ to $473 \times$ when k=$2^{24}$}. Since the performance of bitonic top-$k$ is independent from the data distribution, the speedups over  the normal, uniform and customized distributions are the same. Note, for visualization, we limit the y-axis to [0, 128] in Figure~\ref{fig:all_speedup}. Hence the speedups that are beyond 128 for bitonic top-$k$ are marked as numbers in Figure~\ref{fig:all_speedup}.

\begin{table}[ht]
{
% \scriptsize
% \scriptsize
\small

\begin{tabular}{|c|c|c|c|}
\hline
Dataset       & Abbr. & |V|           & Application domain      \\ \hline
% kmer\_V1r   & KM    & 134,217,728   & Genetics                \\ \hline
ANN\_SIFT1B~\cite{jegou:inria-00514462} & AN    & 536,870,912 & $k$-Nearest Neighbor \\ \hline
ClueWeb09~\cite{nr,clarke2009overview} & CW    & 1,073,741,824 & Sparse Networks \\ \hline
TwitterCOVID-19~\cite{gupta2020global} & TR    & 1,073,741,824 & Social Networks \\ \hline
\end{tabular}
	\caption {Real-world datasets.
% 	\vspace{-.25in}
	}\label{tab:dataset}
	}
\end{table}

\textbf{Real-world datasets} contains three datasets: ANN\_SIFT1B~\cite{jegou:inria-00514462}, ClueWeb09~\cite{CW} and {TwitterCOVID-19}~\cite{gupta2020global}. (i) ANN\_SIFT1B dataset contains 1 billion vectors, each of which is at 128 dimensions and describes an image. We use the first vector from the ANN\_SIFT1B dataset to calculate the euclidean distances between this vector and the 1 billion vectors. Afterward, the distance array is the input vector for top-$k$.
(ii) The ClueWeb09 is a webpage graph which contains 4,780,950,910 webpages and 7,939,635,651 links. We derive the degrees of the webpages and use that as the input vector for top-$k$. {(iii) TwitterCOVID-19}~\cite{gupta2020global} {dataset consists of COVID-fear related scores of the tweets related to the COVID-19 pandemic from 28 January 2020 to 1 January 2021. The original 132 million public twitter posts are duplicated on the vector of 1 billion size to achieve same distribution.}
\textit{Top-$k$ computation can help (i) derive the $k$-NN of a query vector~\cite{jegou:inria-00514462}, (ii) rank the vertices by degree~\cite{doo2014extracting} and {(iii) k least fearful tweets related to the COVID19 pandemic in}~\cite{gupta2020global} {dataset.}}
Since bitonic top-$k$ cannot work on $|V|$ that is not at size of power of 2, and GGKS radix top-$k$ suffers from $|V|\ge2^{31}$, we cut the sizes of (i) and (ii) datasets into 536,870,912 and 1,073,741,824, respectively. 

 \begin{figure}[t]
	\centering
	\includegraphics[width=.47\textwidth]{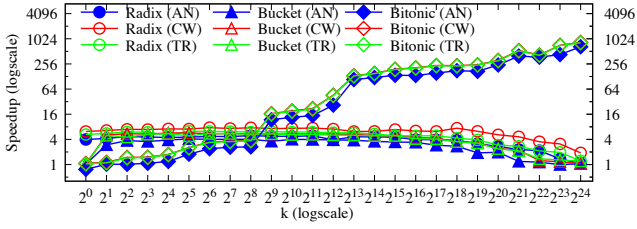}
% 	\vspace{-.15in}
	\caption{{The speedup of} {\topk} {over the-state-of-the-art for a varying $k$ on real-world graph}. 
% 	\vspace{-.2in}
	}
	\label{fig:real_data} 
% 	  \vspace{-.05in}	
\end{figure}
% % We evaluate the real-world datasets for performance analysis of {\topk} over other state-of-the-art tools. Given degree centrality~\cite{doo2014extracting}, and nearest neighbor analysis~\cite{jegou:inria-00514462} can be considered in ranking the vertices in the graph, we use the enlisted datasets in Table~\ref{tab:dataset} for real-world data evaluation. The number of vertices in the dataset  AN and CW are reduced from 1,000,000,000 to 536,870,912 and \fixme{4,780,950,910} to 1,073,741,824 in order to meet the working criteria for bitonic top-$k$.  The score for AN dataset is computed as distance of a node from other nodes, the nodes with lowest score are considered as the most similar elements to the query. The scores of CW datasets are the node degrees in the graph. Top-$k$ on this dataset help rank largely linked webpage in the CW graph. The datasets are availaverview}. 

Figure~\ref{fig:real_data} shows the speedup of {\topk} assisted top-$k$ algorithms over the state-of-the-art projects on the real-world datasets.
In general, for the same top-$k$ algorithm, {\topk} enjoys higher speedups on CW dataset than AN\update{,} because CW is larger. This aligns with our finding in Figure~\ref{fig:FirstTopK_SecondTopKSizePercentage_DiffN}. On average, {\topk} assisted radix, bucket and bitonic top-$k$, respectively, perform 6.7$\times$, 4.6 $\times$ and \update{173.7}$\times$ faster than their corresponding top-$k$ algorithms on {CW}. {AN} dataset observes an average speedup of respectively 4.2$\times$, 3.3$\times$ and \update{127.1}$\times$ over the state-of-the-art top-$k$ algorithms. 
{Similarly, {TR} dataset observes an average speedup of respectively 4.8$\times$, 4.1$\times$ and 170.2$\times$ over the state-of-the-art top-$k$ algorithms.}

\subsection{{\topk} Workload Statistics}
\label{subsec:stat}

 \begin{figure}[h]
	\centering
	\includegraphics[width=.47\textwidth]{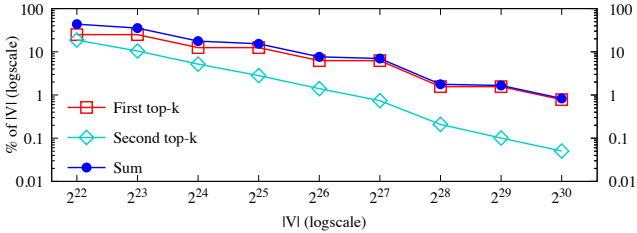}
% 	\vspace{-.1in}
	\caption{Workload dynamics of the first top-$k$, second top-$k$ and their sum with respect to the increase of $|V|$. Here, we set $k=2^{19}$.
% 	\vspace{-.15in}
	}
	\label{fig:FirstTopK_SecondTopKSizePercentage_DiffN} 
	%   \vspace{-.1in}	
\end{figure}

Figure~\ref{fig:FirstTopK_SecondTopKSizePercentage_DiffN} plots the workload dynamics for the first top-$k$, second top-$k$ and their sum with respect to varying sizes of the input vector $|V|$. Particularly, the workloads are the sizes of the delegate vector and the concatenated vector for the first and second top-$k$, respectively. \textit{We observe that the ratio of the delegate vector over $|V|$ decreases significantly when $|V|$ increases, so does that for concatenated vector.} Specifically, the sum of the delegate and concatenated vector sizes is 76.06\% of $|V|$ at $|V|= 2^{22}$ and 0.83\% at $|V|=2^{30}$. \textit{This workload reduction trend demonstrates the scalable nature of {\topk}, that is, {\topk} performance improves when the problem size |V| increases.} 

 \begin{figure}[t]
	\centering
	\includegraphics[width=.47\textwidth]{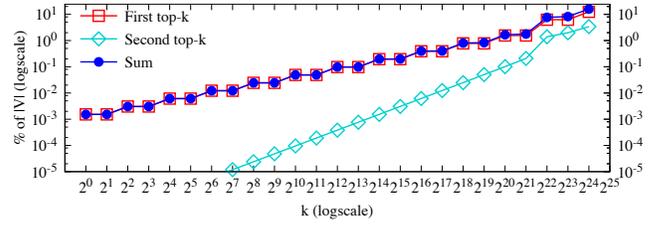}
% 	\vspace{-.1in} %1.png is the relative path of image
	\caption{Workload dynamics of the first top-$k$, second top-$k$ and their sum with respect to the increase of $k$. Here, we set $|V|=2^{30}$.
% 	\fixme{k reduced to $2^{24}$\update{done}}
% 	\vspace{-.15in}
	}

	\label{fig:FirstTopK_SecondTopKSizePercentage} %  used for citing
% 	  \vspace{-.155in}	
\end{figure}

% \begin{tabular}{|c|c|c|c|c|c|c|c|c|c|c|c|c|}

\begin{table*}[ht]
  \centering
{\scriptsize
\setlength\tabcolsep{2.1pt}
\begin{tabular}{
!{\color{black}\vrule width 0.5pt}c!{\color{black}\vrule width 0.5pt}c!{\color{black}\vrule width 0.5pt}c!{\color{black}\vrule width 0.5pt}c!{\color{black}\vrule width 0.5pt}c!{\color{black}\vrule width 0.5pt}c!{\color{black}\vrule width 0.5pt}c!{\color{black}\vrule width 0.5pt}c!{\color{black}\vrule width 0.5pt}c!{\color{black}\vrule width 0.5pt}c!{\color{black}\vrule width 0.5pt}c!{\color{black}\vrule width 0.5pt}c!{\color{black}\vrule width 0.5pt}c!{\color{black}\vrule width 0.5pt}
}
\arrayrulecolor{black}\hline

\multirow{2}{*}{\textbf{\blue{\#GPU (\#Nodes)}}} & \multicolumn{3}{c|}{\textbf{\blue{|V| = $2^{30}$}}}  & \multicolumn{3}{c|}{\textbf{\blue{|V| = $2^{31}$}}}  & \multicolumn{3}{c|}{\textbf{\blue{|V| = $2^{32}$}}}  & \multicolumn{3}{c|}{\textbf{\blue{|V| = $2^{33}$}}}  \\ \cline{2-13} 

 & \begin{tabular}[c]{@{}c@{}}\blue{Communication}\\ \blue{(ms)}\end{tabular} & \begin{tabular}[c]{@{}c@{}}\blue{Reload} \\ \blue{Overhead}\\ \blue{(ms)}\end{tabular} & \begin{tabular}[c]{@{}c@{}}\blue{Total time}\\ \blue{(ms) (speedup)}\end{tabular} & \begin{tabular}[c]{@{}c@{}}\blue{Communication}\\ \blue{(ms)}\end{tabular} & \begin{tabular}[c]{@{}c@{}}\blue{Reload} \\ \blue{Overhead}\\ \blue{(ms)}\end{tabular} & \begin{tabular}[c]{@{}c@{}}\blue{Total time}\\ \blue{(ms) (speedup)}\end{tabular} & \begin{tabular}[c]{@{}c@{}}\blue{Communication}\\\blue{ (ms)}\end{tabular} & \begin{tabular}[c]{@{}c@{}}\blue{Reload} \\ \blue{Overhead}\\ \blue{(ms)}\end{tabular} & \begin{tabular}[c]{@{}c@{}}\blue{Total time}\\ \blue{(ms) (speedup)}\end{tabular} & \begin{tabular}[c]{@{}c@{}}\blue{Communication}\\ \blue{(ms)}\end{tabular} & \begin{tabular}[c]{@{}c@{}}\blue{Reload} \\ \blue{Overhead}\\ \blue{(ms)}\end{tabular} & \begin{tabular}[c]{@{}c@{}}\blue{Total time}\\ \blue{(ms) (speedup)}\end{tabular} \\ \hline
%  \blue{
\blue{1 (1)}& \blue{0} & \blue{0}     & \blue{6.1 (1x)}& \blue{0} &   \blue{373.14}    & \blue{{384.93} (1x)}  & \blue{0} &   \blue{1238.13}    & \blue{{1261.51} (1x)} & \blue{0} &    \blue{2898.54}   & \blue{{2944.99} (1x)}   \\ \hline
\blue{2 (1)} & \blue{0.11} & \blue{0}     & \blue{3.7 (1.6x)}  &  \blue{0.46} &  \blue{0}     & \blue{6.22} \blue{(61.7x)} & \blue{0.06} &   \blue{524.41}    & \blue{536.218 (2.3x)}& \blue{0.08} &     \blue{1586.81}  & \blue{1788.3 (1.7x)}\\ \hline
\blue{4 (1)}& \blue{0.11} & \blue{0}     & \blue{2.5 (2.4x)}  & \blue{0.29} & \blue{0}     & \blue{3.7 (104.0x)} & \blue{0.12} & \blue{0}     & \blue{8.8 (143.3x)}& \blue{0.07} & \blue{1056.02}   & \blue{1067.68 (2.8x)}\\ \hline
\blue{8 (2)}& \blue{0.19} & \blue{0}     & \blue{1.96 (3.1x)} & \blue{0.29} & \blue{0}     & \blue{2.71 (141.7x)}& \blue{0.73} & \blue{0}     & \blue{4.36 (289.2x)}& \blue{1.32} & \blue{0}     & \blue{7.97 (369.4x)}\\ \hline
\blue{16 (4)}  & \blue{0.31} & \blue{0}     & \blue{1.80 (3.4x)} & \blue{0.32} & \blue{0}     & \blue{2.07 (185.9x)}& \blue{0.82} & \blue{0}     & \blue{2.68 (470.5x)}& \blue{1.43} & \blue{0}     & \blue{4.01 (734.2x)}\\ \hline
% }s
\end{tabular}
	\caption {\blue{Scalability of {\topk} with varying $|V|$ and $k=128$.}
% 	\vspace{-.25in}
	}\label{tab:multiGPU}
	}
\end{table*}

Figure~\ref{fig:FirstTopK_SecondTopKSizePercentage} demonstrates the vector size of the first top-$k$ and second top-$k$ in {\topk} assisted radix top-$k$ across different $k$ values. Apparently, {for a given input vector size}, the increase of $k$ will lead to larger vector sizes for both the first and second top-$k$.
As the vector sizes increase to a higher ratio of the input vector, the speedup of {\topk} over the state-of-the-art also decreases. 
Another fact is that the workload of the first top-$k$ dominates the entire workload for {\topk} because $\beta$ delegate will lead to more delegates. Further, $\beta$ delegate and delegate-based filtering together can significantly reduce the workload for the second top-$k$. Particularly, the ratio of the sum of both vectors over the input vector climbs from 0.0015\% to 15.91\% with the increase of $k$.

\begin{figure}[h]
	\centering
\includegraphics[width=0.47\textwidth]{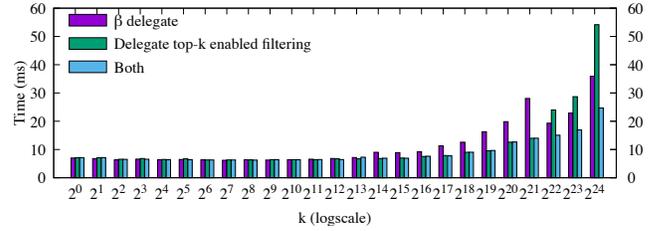}
% \vspace{-.1in}
	\caption{The performance impacts of delegate top-$k$ enabled filtering vs $\beta$ delegate.
% 	\update{Updated}
% 	\fixme{key: $\beta$ delegate}\update{done}
%  	\vspace{-.15in}
	}
	\label{fig:comparison_beta_concat_uniform} 
\end{figure}

Figure~\ref{fig:comparison_beta_concat_uniform} studies the separate and combined effects of delegate top-$k$ based filtering and $\beta$ delegate, given both of them are proposed to reduce the workload for concatenation and second top-$k$. Here, we include delegate vector construction optimization. When $k$ is small, one can observe that delegate top-$k$ enabled filtering yields better performance gains over $\beta$ delegate, $k=2^{20}$ in particular. However, once $k$ becomes bigger, the $\beta$ delegate optimization starts picking up the momentum. Overall, delegate top-$k$ enabled filtering combined with $\beta$ delegate always offers the best performance. Particularly, for \update{$k=2^{24}$}, the time consumption of delegate top-$k$ enabled filtering, $\beta$ delegate and the combined one are \update{54.2} ms, \update{35.9 ms 24.7} ms.

\subsection{{\topk} Scalability}\label{eval:multiGPU}
 
%  \begin{figure}[hbt!]
% 	\centering
% 	\includegraphics[width=.47\textwidth]{./figures/multiGPU.pdf}
% % 	\vspace{-.1in}
% 	\caption{Scalability of {\topk} with varying $|V|$ and $k=128$.
% 	\vspace{-.1in}
% 	\label{fig:multiGPU} 
% 	%   \vspace{-.1in}	
% \end{figure}

% \begin{table}[h]
% {\scriptsize
% \begin{tabular}{|c|c|c|c|c|}
% \hline
% \#GPU & \begin{tabular}[c]{@{}c@{}}$|V| = 2^{30}$\\ ms. (speedup)\end{tabular} & \begin{tabular}[c]{@{}c@{}}$|V| = 2^{31}$\\ ms. (speedup)\end{tabular} & \begin{tabular}[c]{@{}c@{}}$|V| = 2^{32}$\\ ms. (speedup)\end{tabular} & \begin{tabular}[c]{@{}c@{}}$|V| = 2^{33}$\\ ms. (speedup)\end{tabular} \\ \hline
% 1     & 6.1 (1$\times$) & 334.3 (1$\times$)  & 1253.7 (1$\times$)  & 2942 (1$\times$)        \\ \hline
% 2     & 3.7 (1.6$\times$) & 6.1 (54.8$\times$) & 919.9 (1.4$\times$) & 1788.3 (1.6$\times$)                                 \\ \hline
% 4     & 2.5 (2.4$\times$) & 3.7 (90.9$\times$) & 8.8 (142.7$\times$) & 979.5 (3.0$\times$)                                                            \\ \hline
% \end{tabular}
% 	\caption {Scalability of {\topk} with varying $|V|$ and $k=128$.
% 	\vspace{-.25in}
% 	}\label{tab:multiGPU}
% 	}
% \end{table}

{Table~\ref{tab:multiGPU} demonstrates the scalability of {\topk} assisted radix top-$k$ for vector sizes |V| of $2^{30}$ - $2^{33}$ on up to 16 V100 GPUs (4 compute nodes).}  {The table includes the communication overhead among the GPUs, vector reloading overhead, and total time.} Overall, we can observe that {\topk} achieves desirable scalability in various settings. When the partitioned sub-vector can fit in 1 - 16 GPUs in $|V|=2^{30}$, \blue{the speedup goes up to 3.4$\times$ on 16 GPUs. In the remaining columns (|V| $\geq$ $2^{31}$) of the table, we observe superlinear speedup. 
The reason is that when the \# of GPUs is low, we cannot fit all the sub-vectors in GPU before computation. Therefore, {\topk} loads certain partitions during computation. And the total time includes sub-vector loading time.  Whereas, when the GPU number increases to 16, the data can fit in GPUs. 
Thanks to a relatively low communication cost in asynchronous communication for the top-$k$ elements, we observe a maximum communication time of 1.43 ms at 16 GPUs $|V| = 2^{33}$ configuration. Similarly, for a large input vector $|V| = 2^{33}$ and with a single GPU configuration, the reload overload can go up to 2898.54 ms. } 
\blue{Note, we cannot include the multi-GPU settings for the state-of-the-art tools because they do not support multi-GPU features.}

\blue{\subsection{Global Memory Transactions Analysis}}

% \begin{figure}[h]
% 	\centering
% 	\includegraphics[width=.47\textwidth]{./figures/gl_transactions.pdf}
% % 	\vspace{-.15in} 
% 	\caption{\blue{Number of global load and store transactions in different versions in top-k. We test on UD dataset with |V| = $2^{30}$ and k = $2^{7}$.}\fixme{Need to break the y-axis to see all the small data.}
% % 	\vspace{-.2in}
% 	} 
% 	\label{fig:gl_transactions} 
% % 	  \vspace{-.1in}	
% \end{figure}

% \begin{tabular}{|c|c|c|c|c|c|c|}

% \vspace{-.15in}
\begin{table}[hbt!]
  \centering
{
\scriptsize
% \small
\setlength\tabcolsep{3.75pt}
\begin{tabular}{
!{\color{black}\vrule width 0.5pt}
c
!{\color{black}\vrule width 0.5pt}
c
!{\color{black}\vrule width 0.5pt}
c
!{\color{black}\vrule width 0.5pt}
c
!{\color{black}\vrule width 0.5pt}
c
!{\color{black}\vrule width 0.5pt}
c
!{\color{black}\vrule width 0.5pt}
c
!{\color{black}\vrule width 0.5pt}}
\arrayrulecolor{black}\hline
{\color[HTML]{000000} }                                                                                                                     & \multicolumn{2}{c|}{{\color[HTML]{000000} Radix top-k}}       & \multicolumn{2}{c|}{{\color[HTML]{000000} Bucket top-k}}      & \multicolumn{2}{c|}{{\color[HTML]{000000} Bitonic top-k}}                                      \\ \cline{2-7} 
\multirow{-2}{*}{{\color[HTML]{000000} \begin{tabular}[c]{@{}c@{}}\#global memory \\transactions\end{tabular}}} & {\color[HTML]{000000} GGKS~\cite{alabi2012fast}} & {\color[HTML]{000000} Dr.Top-k} & {\color[HTML]{000000} GGKS~\cite{alabi2012fast}} & {\color[HTML]{000000} Dr.Top-k} & {\color[HTML]{000000} Bitonic~\cite{shanbhag2018efficient}} & {\color[HTML]{000000} Dr.Top-k} \\ \hline
{\color[HTML]{000000} \#load ($\times 10^9$)}                                                                                                               & {\color[HTML]{000000} 3.07} & {\color[HTML]{000000} 1.34}     & {\color[HTML]{000000} 4.04} & {\color[HTML]{000000} 1.44}     & {\color[HTML]{000000} 11.45}                                 & {\color[HTML]{000000} 1.35}     \\ \hline
{\color[HTML]{000000} \#store ($\times 10^9$)}                                                                                                              & {\color[HTML]{000000} 2.01} & {\color[HTML]{000000} 0.003}    & {\color[HTML]{000000} 1.36} & {\color[HTML]{000000} 0.003}    & {\color[HTML]{000000} 2.09}  
& {\color[HTML]{000000} 0.007}    \\ \hline
\end{tabular}
	\caption {\blue{Number of global load and store transactions in different versions in top-k. We test on UD dataset with |V| = $2^{30}$ and k = $2^{7}$.}
% 	\vspace{-.25in}
% 	\vspace{-.32in}
	}\label{tab:gl_transactions}
	}
\end{table}

\blue{Table~\ref{tab:gl_transactions} showcases the number of global memory load and store transactions of different versions of top-$k$ on the UD dataset with |V| = $2^{30}$ and k = $2^{7}$. 
We use nvprof~\cite{nvprof} profiler for profiling the results.
% on GGKS Radix, GGKS Bucket top-$k$~\cite{alabi2012fast}, Bitonic~\cite{shanbhag2018efficient} top-$k$, {\topk} assisted radix, bucket and bitonic top-$k$. 
From the table, we observe a reduction of load transactions by 2.3$\times$, 3.1$\times$ and 8.5$\times$, respectively when $\topk$ assists radix, bucket and bitonic top-$k$. Similary, $\topk$ helps reduce the global memory store transaction by 766.8$\times$, 516.9$\times$ and 
%155.7$\times$, 
298.6$\times$,
respectively on radix, bucket and bitonic top-$k$.}

\balance %helps remove a warning if used at the beginning of the last page in the document

\subsection{{\topk} on Different GPUs}
 \begin{figure}[hbt!]
	\centering
	\includegraphics[width=.48\textwidth]{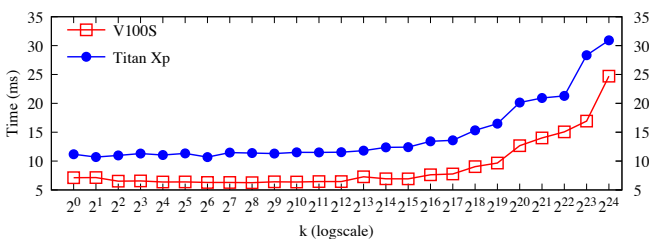} %1.png is the relative path of image
    % \vspace{-.3in}	
	\caption{{\topk} on V100S Vs Titan Xp.
% 	\update{updated}
% 	\fixme{k reduced to $2^{24}$}\update{done}
		  %\vspace{-.15in}	
} %caption is the title of image
	\label{fig:SOK_N_29_SOK_AutoTuneAdaptive_V100_Titan_Comparison} %  used for citing
\end{figure}

Figure~\ref{fig:SOK_N_29_SOK_AutoTuneAdaptive_V100_Titan_Comparison} compares the {\topk} radix top-$k$ on the V100S and Titan Xp GPUs. Clearly, the time consumption patterns of {\topk} on both the GPUs are similar for a range of $k$. Overall, the performance of {\topk} on V100S is better than in Titan Xp by a factor of \update{1.3$\times$ - 1.8$\times$}. This roughly aligns with the ratio of the reported peak throughput difference between V100S~\cite{voltagpu} and Titan Xp~\cite{TitanXp} which are 1,134 GB/s and 547.7 GB/s. 

\subsection{{\topk} vs BMW}\label{eval:BMWVsDrTopk}

\begin{figure}[ht]
	\centering
	\includegraphics[width=.47\textwidth]{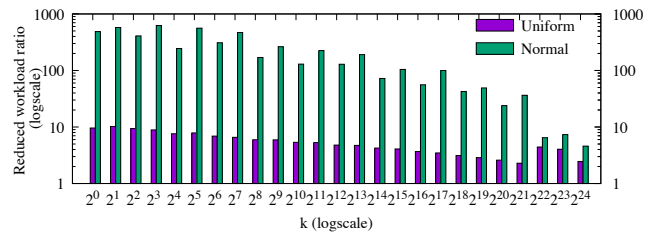}
% 	\vspace{-.15in} 
	\caption{The ratio of fully evaluated workload (after workload reduction) of BMW to that of {\topk}.
% 	\vspace{-.15in}
	} 
	\label{fig:BMW_DrTopk} 
% 	  \vspace{-.05in}	
\end{figure}

Figure~\ref{fig:BMW_DrTopk} presents the $\mathrm{ratio}=\frac{\mathrm{BMW\ workload}}{\mathrm{{\topk}\ workload}}$, where we use the sum the workloads of first and second top-$k$ as {\topk} workload.
Overall, we observe the ratio to be, on average, 212$\times$ in ND and 6$\times$ in UD, which suggests that {\topk} reduces $212\times$ and $6\times$ more workload than BMW. The reason is that BMW still works on each regular item even after deriving the delegate while {\topk} uses a delegate to skip an entire subrange directly.

\section{Conclusion}\label{sec:conclusion}
We introduce {\topk} with three contributions:
First and foremost, {\topk} introduces a comprehensive delegate-centric concept to help tremendously reduce the workload for top-$k$ computations. Second, we introduce a practical way to partition the input vector into proper sized subranges with theoretical support. Finally, we deploy our project atop distributed GPUs to handle extreme large input vectors along with various system optimizations. Taken together, {\topk} assisted top-$k$ algorithms constantly outperform the state-of-the-art.

\section*{Acknowledgement}
This  work  was  in  part  supported  by NSF CRII Award No. 2000722, CAREER Award No. 2046102, Australia Research Council (ARC) Discovery Project DP210101984, and the Exascale Computing  Project  (17-SC-20-SC),  a  collaborative  effort  of  the  U.S. Department of Energy Office of Science and the National Nuclear Security Administration (NNSA). Any opinions, findings and conclusions, or recommendations expressed in this material are those of the authors and do not necessarily reflect the views of DOE, NNSA, ARC, or NSF.

\bibliographystyle{acm}
% \bibliography{sample-bibliography}
\bibliography{sampleK}

\end{document}